\documentclass[11pt,a4paper]{article}
\pdfoutput=1
\usepackage{geometry}   
\usepackage[toc,page]{appendix}
\usepackage{amsfonts,amssymb,amsmath,mathrsfs}
\usepackage{graphicx}
\usepackage{hyperref}
\usepackage[bbgreekl]{mathbbol}
\usepackage{color,xcolor}
\usepackage{comment}
\usepackage{cancel}
\usepackage[utf8]{inputenc}

\usepackage[textsize=footnotesize,textwidth=1.5cm]{todonotes}
\usepackage{caption}

\numberwithin{equation}{section}

\def\){\right)}
\def\({\left( }
\def\]{\right] }
\def\[{\left[ }

\def\NO{\nonumber}

\newcommand{\be}{\begin{equation}}
\newcommand{\ee}{\end{equation}}

\def\bea{\begin{eqnarray}}
\def\eea{\end{eqnarray}}

\def\bal#1\eal{\begin{align}#1\end{align}}

\def\bald{\begin{aligned}}
\def\eald{\end{aligned}}

\def\bsub{\begin{subequations}}
\def\esub{\end{subequations}}

\def\beqx{\begin{displaymath}}
\def\eeqx{\end{displaymath}}

\newcommand{\bmat}{\left(\begin{array}}
\newcommand{\emat}{\end{array}\right)}

\def\a{\alpha}
\def\b{\beta}
\def\c{\chi}
\def\d{\delta}
\def\e{\epsilon}

\def\j{\psi}
\def\k{\kappa}
\def\l{\lambda}
\def\m{\mu}
\def\n{\nu}
\def\o{\omega}
    
\def\p{\pi}

\def\r{\rho}
\def\s{\sigma}
\def\t{\tau}
\def\x{\xi}

\def\F{\Phi}
\def\G{\Gamma}

\def\L{\Lambda}
\def\O{\Omega}
    
\def\P{\Pi}

\def\S{\Sigma}

\def\ve{\varepsilon}

\def\cb{{\cal B}}
\def\cc{{\cal C}}

\def\ce{{\cal E}}
\def\cf{{\cal F}}

\def\ci{{\cal I}}
\def\cj{{\cal J}}
\def\ck{{\cal K}}
\def\cl{{\cal L}}
\def\cm{{\cal M}}

\def\co{{\cal O}}
\def\cp{{\cal P}}
\def\cq{{\cal Q}}
\def\car{{\cal R}}
\def\cs{{\cal S}}
\def\ct{{\cal T}}

\def\cv{{\cal V}}
\def\cw{{\cal W}}

\def\bb#1{\ensuremath{\mathbb{#1}}} 

\def\bo{{\raise-.3ex\hbox{\large$\Box$}}}               
\def\pa{\partial}                                       
\def\face{{\raise.2ex\hbox{$\displaystyle \bigodot$}\mskip-2.2mu \llap {$\ddot
        \smile$}}}                                   
\def\>{\rangle}                                      
\def\<{\langle}                                      

\def\tx#1{\text{#1}}
\def\sbtx#1{{}_{\rm #1}}                           
\newcommand{\sub}[1]{\phantom{}_{(#1)}\phantom{}}    
\def\wt#1{\widetilde{#1}}                            
\def\Hat#1{\widehat{#1}}                             
\def\lbar#1{\ensuremath{\overline{#1}}}              
\def\leftrightarrowfill{$\mathsurround=0pt \mathord\leftarrow \mkern-6mu
        \cleaders\hbox{$\mkern-2mu \mathord- \mkern-2mu$}\hfill
        \mkern-6mu \mathord\rightarrow$}        
\def\dvec#1{\vbox{\ialign{##\crcr
        \leftrightarrowfill\crcr\noalign{\kern-1pt\nointerlineskip}
        $\hfil\displaystyle{#1}\hfil$\crcr}}}           
\def\tr{{\rm tr \,}}                                    

\def\-{\hphantom{-}}

\begin{document}

\title{\vspace{-1.0in}
\begin{minipage}[t]{\textwidth}
 \begin{flushright}
 {\small{KIAS-P20010}}\\\vskip2.0in
 \end{flushright}
\end{minipage}\\ 
\vspace{1.0in}{\bf Counterterms, Kounterterms, and the variational problem in AdS gravity}}

\author{Giorgos Anastasiou$^{1}$\footnote{georgios.anastasiou@pucv.cl}\thinspace , 
Olivera Miskovic$^{1}$\thanks{olivera.miskovic@pucv.cl}\thinspace , 
Rodrigo Olea$^{2}$\thanks{rodrigo.olea@unab.cl}\thinspace ,   \and and Ioannis Papadimitriou$^{3}$\thanks{ioannis@kias.re.kr} \bigskip \\
{\small {\ $^{1}$Instituto de F\'{\i}sica, Pontificia Universidad Cat\'{o}lica de Valpara\'{\i}so,}}\\
{\small {Casilla 4059, Valpara\'{\i}so, Chile}}\\
{\small {$^{2}$}Departamento de Ciencias F\'{\i}sicas, Universidad Andres
Bello, Sazi\'{e} 2212, Piso 7, Santiago, Chile}\\
{\small {$^{3}$}School of Physics, Korea Institute for Advanced Study, Seoul 02455, Korea}}

\date{}

\maketitle
\thispagestyle{empty}

\begin{abstract}
We show that the Kounterterms for pure AdS gravity in arbitrary even dimensions coincide with the boundary counterterms obtained through holographic renormalization if and only if the boundary Weyl tensor vanishes. In particular, the Kounterterms lead to a well posed variational problem for generic asymptotically locally AdS manifolds only in four dimensions. We determine the exact form of the counterterms for conformally flat boundaries and demonstrate that, in even dimensions, the Kounterterms take exactly the same form. This agreement can be understood as a consequence of Anderson's theorem for the renormalized volume of conformally compact Einstein 4-manifolds and its higher dimensional generalizations by Albin and Chang, Qing and Yang. For odd dimensional asymptotically locally AdS manifolds with a conformally flat boundary, the Kounterterms coincide with the boundary counterterms except for the logarithmic divergence associated with the holographic conformal anomaly, and finite local terms.  
\end{abstract}

\newpage
\tableofcontents
\setcounter{page}{1}

\section{Introduction}
\label{sec:intro}

Hyperbolic manifolds and their pseudo Riemannian cousins, de Sitter (dS) and anti de Sitter (AdS) space, arise in several contexts in physics and mathematics. Cosmic inflation in the early universe and the dark energy dominated expansion at late times are well approximated by dS space, while AdS space is the arena of most holographic dualities. In mathematics, the work of Fefferman and Graham \cite{fefferman1985mathematical} relates hyperbolic geometry with the study of conformal invariants, whereas Hyperbolic 3-manifolds provide deep connections between number theory, topology and geometry.

The focus of the present work are general asymptotically locally AdS (AlAdS) manifolds, known as conformally compact Einstein or Poincar\'e-Einstein manifolds in the mathematics literature. These are Riemannian or pseudo Riemannian solutions of Einstein's equations with a negative cosmological constant, but the aspects we will discuss here apply also to solutions with a positive cosmological constant, that is  asymptotically locally dS manifolds. A common property of all such manifolds is that they have an infinite volume and a compact conformal boundary. 

Conformal geometry on the boundary of AlAdS manifolds can be studied through hyperbolic geometry in the interior \cite{fefferman1985mathematical}. In particular, certain quantities obtained from the bulk geometry, such as the renormalized volume in even dimensions and Branson's $Q$-curvature in odd dimensions, compute boundary conformal invariants \cite{graham2001scattering,FG-Q-curvature}. A result of direct relevance to our  analysis was proved by Anderson for the case of four dimensional AlAdS manifolds \cite{Anderson_l2} and was generalized to higher even dimensions by Albin \cite{albin2005renormalizing} and Chang, Qing and Yang \cite{chang2005renormalized}. Anderson's result concerns the renormalized volume of AlAdS$_4$ manifolds and is summarized in the formula
\be\label{AndersonTh}
\frac{1}{8(2\p)^2}\int_{\cm_4} |W|^2+\frac{3}{(2\p)^2}V\sbtx{ren}(\cm_4)=\c(\cm_4 ),
\ee
where $W$ is the Weyl tensor of the bulk metric, $\c(\cm_4)$ is the Euler characteristic of $\cm_4$, and $V\sbtx{ren}(\cm_4)$ is the renormalized volume. It is instructive to compare this with the generalized Chern-Gauss-Bonnet theorem for manifolds with boundary \cite{Chern1945} (see \eqref{CGB-theorem} below)
\be\label{CGB-theorem-intro}
\int_{\cm_4}\O+\int_{\pa\cm_4}\P=\c(\cm_4),
\ee
where $\O$ is the Pfaffian of the bulk Riemann tensor, i.e. the Euler-Poincar\'e density, and the Chern form, $\P$, satisfies $-\tx d\P=\O$. Since the Weyl tensor is the traceless part of the Riemann tensor, it follows that for any Einstein manifold we have, schematically,   
\be
\int_{\cm_4}\O\sim \int_{\cm_4} |W|^2+\L\, V(\cm_4),
\ee
where $\L$ is the cosmological constant and $V(\cm_4)$ is the volume of $\cm_4$, defined with some regulator. 

Comparing \eqref{AndersonTh} and \eqref{CGB-theorem-intro}, we see that the content of Anderson's theorem is that the renormalized volume of four dimensional AlAdS manifolds is (again schematically) given by
\be\label{Vren}
V\sbtx{ren}(\cm_4)\sim V(\cm_4)+\L^{-1}\int_{\pa\cm_4}\P.
\ee 
In particular, the Chern form associated with the Pfaffian of the bulk Riemann tensor renormalizes the volume of AlAdS$_4$ manifolds. As we will see in the subsequent analysis, this conclusion hinges crucially on the fact that the integral of the square of the Weyl tensor over $\cm_4$ is finite. This does not hold for higher even dimensional AlAdS manifolds. Of course, neither the Chern-Gauss-Bonnet theorem nor Anderson's theorem can be extended to odd dimensional AlAdS manifolds.   

In the context of the AdS/CFT correspondence \cite{Maldacena:1997re}, the renormalized volume of even dimensional AlAdS manifolds is interpreted as the partition function of the dual conformal field theory (CFT), while the $Q$-curvature of odd dimensional AlAdS manifolds corresponds to the conformal anomaly of the dual CFT \cite{Henningson:1998gx}. Both these quantities can be computed through a systematic procedure known as holographic renormalization \cite{Henningson:1998gx,deHaro:2000xn,Papadimitriou:2004ap}. More generally, holographic renormalization computes the boundary term required in order to formulate the variational problem in terms of conformal equivalence classes of boundary data, rather than conformal representatives \cite{Papadimitriou:2005ii,Papadimitriou:2010as}. This is necessary for a well posed variational problem on a conformal boundary and is related with the ellipticity of boundary conditions at the quantum level \cite{Witten:2018lgb}. 

Borrowing terminology from the AdS/CFT context, we will refer to the boundary term that allows the variational problem on AlAdS manifolds to be formulated in terms of conformal classes on the boundary as `boundary counterterms'. Several properties of this boundary term are universal, yet often overlooked. Firstly, it must be covariant and local, i.e. analytic in field space and polynomial in boundary derivatives. In any situation where these two properties cannot be maintained simultaneously, locality is given priority at the expense of covariance. In the holographic context, such situations indicate the presence of an anomaly in the dual field theory. Notice that locality of the boundary counterterms is related with the compactness of the boundary. The boundary of any AlAdS manifold is compact, but there are instances where this property is not manifest, leading occasionally to the erroneous conclusion that non-local counterterms are required. An example is the AdS$_d$ slicing of AdS$_{d+1}$. A less trivial one is the Janus solution of type IIB supergravity \cite{Bak:2003jk}. Once the correct conformal compactification is identified, however, even in such cases the boundary can be shown to be compact \cite{Papadimitriou:2004rz} and the corresponding boundary term local.      

Another general property of the boundary counterterms for AlAdS manifolds is that their divergent part is unique. The only ambiguity is the possibility of adding an arbitrary linear combination of boundary conformal invariants, which contribute finite terms only. Due to its interpretation in the the context of the AdS/CFT correspondence, this freedom is referred to as `renormalization scheme dependence'. The uniqueness of the divergent part of the boundary counterterms, however, implies that {\em any} boundary term that renders the variational problem on AlAdS manifolds well posed must coincide with the boundary counterterms, possibly up to finite terms. In combination with Anderson's theorem, this suggests that the boundary counterterms for Einstein-Hilbert gravity on four dimensional AlAdS manifolds should be given by the Chern form, $\P$. We show that this is indeed the case, and  generalize this statement to AlAdS manifolds of arbitrary even dimension.       

The observation that the Chern form renormalizes the volume of AlAdS$_4$ manifolds was also the inspiration for the Kounterterms, first proposed for even dimensional AlAdS manifolds in  \cite{Olea:2005gb} and later generalized to odd dimensions in \cite{Olea:2006vd}. In even dimensions, the Kounterterms are nothing but the pullback of the Chern form, $\P$, on the boundary $\pa\cm$. They are a polynomial in the extrinsic curvature of the induced metric on $\pa\cm$, which corresponds to the pullback of the bulk connection one-form onto $\pa\cm$. The Kounterterms for odd dimensions are also a polynomial in the extrinsic curvature of the boundary, but they are not related with the Chern form in that case. 

Given that the Kounterterms are expressed in terms of the extrinsic curvature, while the counterterms are a polynomial in the intrinsic curvature of the induced metric on $\pa\cm$, a direct comparison seems impossible. However, the variational problem on AlAdS manifolds can be formulated only within the space of metrics that are asymptotically Einstein, which implies that the extrinsic curvature and the induced metric on $\pa\cm$ are asymptotically related. Using this on-shell relation, the Kounterterms can be rewritten entirely in terms of the intrinsic curvature of the induced metric, permitting a direct comparison with the counterterms. Since these are unique, the Kounterterms lead to a well posed variational problem only if the two coincide, at least up to finite local terms.  

Comparing the Kounterterms with the boundary counterterms for generic AlAdS manifolds in dimensions three to seven, we show that a necessary condition for agreement is that the Weyl tensor of the boundary metric be zero. This is automatically satisfied for AlAdS$_4$ manifolds, since the Weyl tensor in three dimensions vanishes identically, and reflects the fact, pointed out above, that the integral of the square of the Weyl tensor is finite in four dimensions, but not in higher dimensions. 
For odd dimensional AlAdS manifolds, a second necessary condition for the Kounterterms to agree with the boundary counterterms is that the Euler characteristic of the boundary also vanishes.
With the exception of AlAdS$_4$ manifolds, therefore, the Kounterterms do not lead to a well posed variational problem for generic AlAdS manifolds.      

We determine the general form of both the counterterms and Kounterterms for AlAdS manifolds with a conformally flat boundary of arbitrary dimension, and we demonstrate that the vanishing of the boundary Weyl tensor, as well as of the boundary Euler characteristic in the case of odd dimensions, are also sufficient conditions for the Kounterterms to coincide with the boundary counterterms. These conditions are summarized in table~\ref{conditions}.

\begin{minipage}{6in}
\vskip0.5cm
\begin{center}
\begin{tabular}{|c|c|}
\hline
$d=$ dim$(\pa\cm)$ & Conditions\\
\hline 
2 & Euler$(\pa\cm)=0$ \\
3 & --\\
even $>2$ & Weyl$(\pa\cm)=0$ \& Euler$(\pa\cm)=0$\\
odd $>3$& Weyl$(\pa\cm)=0$ \\
\hline
\end{tabular} 
\captionof{table}{Necessary and sufficient conditions for the validity of the boundary Kounterterms. Except in four dimensions ($d=3$), the Kounterterms regulate the AdS variational problem if and only if the Weyl tensor of the boundary is zero (odd $d>3$), or both the Weyl tensor and Euler density of the boundary vanish (even $d$).} 
\label{conditions}
\end{center}
\vskip0.5cm
\end{minipage}

Finally, it is worth emphasizing that the boundary Kounterterms do not correspond to an alternative renormalization scheme in the AdS/CFT sense. Unless the conditions in table~\ref{conditions} are met, the Kounterterms do not regulate the variational problem, nor do they remove the long distance divergences of the on-shell action. Moreover, whenever the conditions in table~\ref{conditions} are satisfied, we find that the Kounterterms correspond to a minimal subtraction scheme, i.e. they coincide with the boundary counterterms without any additional finite local contributions.      

This paper is organized as follows. In section \ref{sec:hr} we provide a self contained overview of the dilatation operator method of holographic renormalization for pure AdS gravity. We emphasize the uniqueness of the divergent boundary counterterms and their role in the regularization of the variational problem on asymptotically locally AdS manifolds. The ambiguity corresponding to the choice of finite local counterterms is also discussed in detail. In section \ref{sec:Kounterterms} we review the Chern-Gauss-Bonnet theorem for manifolds with boundary and we explain how it naturally leads to the construction of the Kounterterms for AdS gravity. Expressing the Kounterterms in terms of the the intrinsic curvature of the boundary in dimensions three to seven, we compare them with the boundary counterterms and show that there is no agreement unless the boundary Weyl tensor vanishes. Section \ref{sec:conflat} focuses on asymptotically locally AdS manifolds with a vanishing boundary Weyl tensor. We determine the boundary counterterms in arbitrary dimension for such manifolds and show that in even dimensions they coincide with the Kounterterms, while in odd dimensions we pinpoint the difference. We conclude in section \ref{disc} with a brief discussion. A number of auxiliary technical results are collected in two appendices.

\section{Boundary counterterms from a variational principle}
\label{sec:hr}

A well posed variational principle on a non compact manifold requires the addition of suitable boundary terms. In this section we review the connection between the variational problem for pure Einstein-Hilbert gravity in asymptotically locally anti de Sitter (AlAdS) backgrounds and the local boundary counterterms required to render it well posed. In particular, we show that the boundary counterterms satisfy the radial Hamilton-Jacobi equation, which can be most efficiently solved iteratively using the dilatation operator method \cite{Papadimitriou:2004ap}.

\subsection{Asymptotically locally AdS manifolds}

A non compact (pseudo) Riemannian manifold is said to be AlAdS if it is a conformally compact Einstein manifold (also known as a Poincar\'e-Einstein manifold), which is defined as follows \cite{fefferman1985mathematical,Penrose:1986ca,Graham:1999jg,Skenderis:2002wp,Anderson:2004yi,chang2005renormalized}. If $\cm$ denotes the interior of a $d+1$ dimensional compact manifold $\lbar\cm$ with boundary $\pa\cm$, then a (pseudo) Riemannian metric $g$ on $\cm$ is said to be conformally compact if there exists a smooth and non-negative function $\O$ on $\lbar\cm$, such that $\O(\pa\cm)=0$, $\tx d\O(\pa\cm)\neq 0$, and $\wt g=\O^2 g$ extends smoothly to a non-degenerate metric on $\lbar\cm$, i.e. $g$ has a second order pole at the boundary. If it exists, the defining function $\O$ is not unique and hence the conformal compactification is not unique. In particular, the metric $g$ on $\cm$ induces only a conformal class $[g_{(0)}]$ of boundary metrics $g_{(0)}=\left.\wt g\right|_{\pa\cm}$. In the vicinity of the conformal boundary, the Ricci and Riemann tensors of conformally compact manifolds behave respectively as 
\bal\label{ConfComp}
R_{\m\n}[g]=&\;-d\; |\tx d\O|^2_{\wt g}\;g_{\m\n}+\co(\O^{-1}),\NO\\
R_{\m\n\r\s}[g]=&\;|\tx d\O|^2_{\wt g}\(g_{\m\s}g_{\n\r}-g_{\m\r}g_{\n\s}\)+\co\(\O^{-3}\),
\eal
where
\be
|\tx d\O|^2_{\wt g}\equiv\wt g^{\m\n}\pa_\m\O\pa_\n\O=\co(\O^0).
\ee

The asymptotic form \eqref{ConfComp} of the Riemann tensor implies that the corresponding Weyl tensor is asymptotically subleading relative to $R_{\m\n\r\s}[g]$. This follows from the fact that the Weyl tensor
\be\label{BWeyl}
W_{\m\n\r\s}\equiv R_{\m\n\r\s}+g_{\m\s}P_{\n\r}+g_{\n\r}P_{\m\s}-g_{\m\r}P_{\n\s}-g_{\n\s}P_{\m\r},
\ee
where the Schouten tensor $P_{\m\n}$ in $d+1$ dimensions is defined as  
\be\label{BSchouten}
P_{\m\n}=\frac{1}{d-1}\left(R_{\m\n}-\frac{1}{2d}R g_{\m\n}\right),
\ee
transforms homogeneously under local Weyl rescalings of $g_{\m\n}$. Namely,  
\be
W_{\m\n\r\s}[g]=\O^{-2}W_{\m\n\r\s}[\wt g]=\co(\O^{-2}), 
\ee
while the leading asymptotic behavior of $R_{\m\n\r\s}[g]$ as $\O\to 0$ is $\co(\O^{-4})$.

An AlAdS manifold is a conformally compact manifold that is also Einstein, i.e. it satisfies Einstein's equations with a negative cosmological constant\footnote{More generally, AlAdS manifolds are solutions of Einstein's equations with a matter stress tensor that is asymptotically subleading relative to the cosmological constant term.}
\be\label{einstein}
R_{\m\n}-\frac{1}{2}R g_{\m\n}+\L g_{\m\n}=0,
\ee
where 
\be\label{CosmConst}
\L=-\frac{d(d-1)}{2\ell^{2}},
\ee
and $\ell$ is the AdS radius. In combination with the asymptotic behavior of the Ricci tensor in \eqref{ConfComp}, Einstein's equations imply that   
\be\label{AlAdS}
|\tx d\O|^2_{\wt g}=\frac{1}{\ell^2}.
\ee

Using the Gaussian normal coordinate $\r$ emanating from the conformal boundary $\pa\cm$ as the asymptotic radial coordinate on $\lbar\cm$, the non-degenerate metric $\wt g$ takes the form 
\be\label{FG-gauge}
\wt g=\tx d\r^2+g_\r,\qquad g_\r= g_{(0)}+\co(\r),
\ee
where $g_{(0)}$ is a non-degenerate metric on $\pa\cm$. For a defining function that only depends on the radial coordinate, i.e. $\O=\O(\r)$, the condition \eqref{AlAdS} determines  
\be
\O=\frac\r\ell.
\ee
It follows that the AlAdS metric $g$ admits the asymptotic (Fefferman-Graham) form  \cite{fefferman1985mathematical,Graham:1999jg,Skenderis:2002wp,Anderson:2004yi}
\be
g=\frac{\ell^2}{\r^2}\big(\tx d\r^2+g_\r\big)=\frac{\ell^2}{\r^2}\big(\tx d\r^2+g_{(0)}+\co(\r)\big).
\ee
In the subsequent analysis, it will be useful to introduce the non compact radial coordinate $r=-\ell\log(\r/\ell)$ so that 
\be\label{FG-canonical}
g=\tx dr^2+h_{ij}(r,x)\tx dx^i\tx dx^j,\qquad h_{ij}(r,x)=e^{2r/\ell}\big(g_{(0)ij}(x)+\co(e^{-r/\ell})\big),\qquad i,j=1,\ldots d.
\ee

\subsubsection*{Penrose-Brown-Henneaux diffeomorphisms} 

Since an AlAdS metric on $\cm$ induces only a conformal class of metrics on $\pa\cm$, any specific choice of radial coordinate, such as the one in \eqref{FG-canonical}, is only defined up a residual bulk coordinate transformation that preserves the asymptotic form of the metric but acts non trivially within the conformal class of boundary metrics through a Weyl transformation, namely   
\be\label{boundaryWeyl}
g_{ij(0)}(x)\to e^{2\s(x)/\ell}g_{ij(0)}(x).
\ee
These residual bulk coordinate transformations are known as Penrose-Brown-Henneaux (PBH) diffeomorphisms \cite{Imbimbo:1999bj} and take the form\footnote{Of course, arbitrary transverse diffeomorphisms of the form $x^i\to x'^i=f^i(x)$ also preserve the form \eqref{FG-canonical} of the metric, but do not act on the radial coordinate.} 
\be\label{PBH}
r\to r'= r+\s(x),\qquad x^i\to x'^i=x^i+\frac{\ell}{2}e^{-2r/\ell}g^{ij}_{(0)}(x)\pa_j\s(x)+\co(e^{-3r/\ell}),
\ee
where $\s(x)$ is an arbitrary function of the transverse coordinates. As we will see in the next subsection, these residual bulk diffeomorphisms play a crucial role in the formulation of a well posed variational problem on AlAdS manifolds. 

\subsection{The variational problem in terms of conformal classes} 

We have seen that an AlAdS metric on $\cm$ induces only a conformal class of metrics on the conformal boundary, $\pa\cm$, and so the variational problem on $\cm$ must be formulated in terms of the conformal class, $[g_{(0)}]$, instead of the conformal representative $g_{(0)}$. In particular, the variational problem is well posed provided the on-shell action is a class function on $\pa\cm$ \cite{Papadimitriou:2005ii,Papadimitriou:2010as}.\footnote{The on-shell action cannot be rendered a class function for even $d$ due to the conformal anomaly \cite{Henningson:1998gx}, but it can still furnish a representation of the Abelian group of Weyl transformations.} The relation between boundary Weyl transformations and the bulk diffeomorphisms \eqref{PBH} maps any boundary class function to a function of the bulk metric that is invariant under radial translations. Requiring the on-shell action, evaluated with a radial cutoff, to be independent of the cutoff position, determines the boundary counterterms, up to a finite set of local conformal invariants on the boundary. 

In this paper we focus exclusively on the variational problem for the Einstein-Hilbert action
\be\label{action}
S=\frac{1}{2\k^2}\Big(\int_{\cm}\tx d^{d+1}x\,\sqrt{-g}(
R-2\Lambda ) +\int_{\pa\cm}\tx d^dx\,\sqrt{-h}\;2K\Big),
\ee
where $\k^2=8\p G$ is the gravitational constant in $d+1$ dimensions, the cosmological constant $\L$ is given in \eqref{CosmConst}, and the surface term involving the trace, $K$, of the extrinsic curvature of $\pa\cm$ is the standard Gibbons-Hawking term \cite{Gibbons:1976ue}. The field equations following from this action are Einstein's equations \eqref{einstein}, which admit AlAdS solutions. 

In order to formulate the variational problem, it is necessary to regularize $\cm$ by introducing a radial cutoff surface infinitesimally away from $\O=0$ and consider instead $\pa\cm_\e =\O^{-1}(\e)$, where $\e$ is a small and positive number. This amounts to introducing an upper bound $r_c$ on the radial coordinate $r$ in \eqref{FG-canonical}, which explicitly breaks the PBH diffeomorphisms \eqref{PBH}. These diffeomorphisms imply that moving the position of the radial cutoff $r_c$ is equivalent to changing the conformal representative of the conformal class of boundary metrics $[g_{(0)}]$. Hence, rendering the variational problem independent of the position of the radial cutoff is equivalent to a variational principle in terms of conformal classes of boundary metrics. Moreover, if the variational problem is independent of the location of the radial cutoff, the on-shell action remains finite as the cutoff is removed. We will now show that the regularized variational problem can be rendered independent of the radial cutoff by adding a suitable boundary term.  

\subsubsection*{General variations}
The general variation of the action \eqref{action} on the regularized manifold $\cm_{r_c}$ takes the form
\be\label{variation}
\d S_{\rm reg}=\frac{1}{2\k^2}\int_{\cm_{r_c}}\tx d^{d+1}x\,\sqrt{-g}\Big(R_{\m\n}-\frac{1}{2}R g_{\m\n}+\L g_{\m\n}\Big)\d g^{\m\n}+\int_{\pa\cm_{r_c}}\hskip-0.2cm\tx d^dx\,\p^{ij}\d h_{ij},
\ee
where $h_{ij}$ is the induced metric on the regularized boundary $\pa\cm_{r_c}$ and $\p^{ij}$ is its conjugate canonical momentum in the Hamiltonian formulation of the dynamics where the radial coordinate $r$ plays the role of Hamiltonian `time'. Notice that the $g_{rr}$ and $g_{ri}$ components of the metric do not enter in the variational problem. These components correspond respectively to the lapse and shift functions in the radial ADM formalism \cite{Arnowitt:1960es}, which are non dynamical Lagrange multipliers with vanishing canonical momenta. In the gauge \eqref{FG-canonical}, corresponding to setting the lapse and shift functions respectively to 1 and 0, the canonical momentum of the induced metric $h_{ij}$ takes the form
\be\label{momentum}
\p^{ij}=\frac{1}{2\k^2}\sqrt{-h}\(Kh^{ij}-K^{ij}\),
\ee
where $K_{ij}=\frac12 \dot h_{ij}$ and $K=h^{ij}K_{ij}$ denote respectively the extrinsic curvature of $\pa\cm_{r_c}$ in $\lbar\cm_{r_c}$ and its trace, with the dot in $\dot h_{ij}$ indicating a total derivative with respect to the radial coordinate $r$. Notice that, up to the volume element, the canonical momentum \eqref{momentum} coincides with the quasilocal Brown-York stress tensor \cite{Brown:1992br}
\be
T^{ij}_{\rm BY}=\frac{1}{2\k^2}\(Kh^{ij}-K^{ij}\).
\ee
The variational principle \eqref{variation} demonstrates that the variational problem on AlAdS manifolds is inherently related to a radial Hamiltonian formulation of the dynamics.

\subsubsection*{Diffeomorphisms and variations of the radial cutoff} 

The Lagrangian of a diffeomorphism invariant theory transforms as a tensor density under diffeomorphisms. Namely, under an infinitesimal coordinate transformation, $x^\m\to x^\m+\x^\m$, the regularized action \eqref{action} transforms as
\be\label{diff-variation}
\d_{\x} S_{\rm reg}=\int_{\pa\cm_{r_c}}\hskip-0.2cm\tx d^dx\,\x^r \mathscr{L},
\ee
where in the gauge \eqref{FG-canonical} the radial Lagrangian density, $\mathscr{L}$,  takes the form
\be\label{rLagrangian}
\mathscr{L}=\frac{1}{2\k^2}\sqrt{-h}\big(\car[h]-2\L+K^2-K^i_jK^j_i\big), 
\ee
and $\car[h]$ denotes the Ricci curvature of the induced metric $h_{ij}$. An alternative way to derive \eqref{diff-variation} is to use the transformation of the metric $g_{\m\n}$ on $\cm_{r_c}$ and of the induced metric $h_{ij}$ on $\pa\cm_{r_c}$ under diffeomorphisms, respectively $\d_\x g_{\m\n}=\nabla_\m\x_\n+\nabla_\n\x_\m$ and $\d_\x h_{ij}=D_i\x_j+D_j\x_i+2K_{ij}\x^r$, in the general variation of the regularized action in \eqref{variation}. Throughout this paper, $\nabla_\m$ denotes the covariant derivative with respect to the bulk metric $g_{\m\n}$, while $D_i$ stands for the covariant derivative with respect to the induced metric $h_{ij}$. 

The transformation \eqref{diff-variation} reflects the fact that diffeomorphisms with $\x^r\neq 0$ are not a symmetry of the regularized theory. These are precisely the PBH diffeomorphisms \eqref{PBH}, which correspond to a translation of the radial cutoff $r_c$ and induce a Weyl transformation on the boundary metric $g_{(0)}$. Specifically, \eqref{diff-variation} implies that under an infinitesimal PBH transformation
\be
\x^r=\d\s(x),\qquad
\x^i=\frac{\ell}{2}e^{-2r/\ell}g^{ij}_{(0)}(x)\pa_j\d\s(x)+\co(e^{-3r/\ell}),
\ee
the regularized action transforms as
\be\label{}
\d_\s S_{\rm reg}=\frac{1}{\k^2}\int_{\pa\cm_{r_c}}\hskip-0.2cm\tx d^dx\,\sqrt{-h}\,\d\s\big(\car[h]-2\L+K^2-K^i_jK^j_i\big)\neq 0.
\ee
As it stands, therefore, the variational problem on the cutoff surface depends explicitly on the conformal representative of the conformal class of boundary metrics $g_{(0)}$. 

Radial diffeomorphisms can be restored as a symmetry of the theory on a non compact manifold by imposing suitable boundary conditions and adding the corresponding boundary terms. The relevant boundary condition in this case is that the metric on $\cm$ be AlAdS, which projects the field configurations onto the space of asymptotic solutions of the field equations. As a consequence, the canonical variables $h_{ij}$ and $\p^{ij}$, or equivalently $h_{ij}$ and $K_{ij}$, are asymptotically on-shell and are therefore not independent. The unique asymptotic relation $K_{ij}[h]$ between the variables $h_{ij}$ and $K_{ij}$ that any AlAdS metric obeys is the key to determining the boundary terms necessary to restore the radial diffeomorphisms as a symmetry of the theory on $\cm$. This relation, however, also means that the relevant boundary term can be equivalently expressed in terms of $h_{ij}$ or $K_{ij}$. Indeed, using the first Gauss-Codazzi equation in \eqref{GC}, the transformation \eqref{diff-variation} of the regularized action can be written on-shell in different ways:  
\be\label{diff-variation-onshell}
\left.\d_\x S_{\rm reg}\right|\sbtx{on-shell}=\frac{1}{\k^2}\int_{\pa\cm_{r_c}}\hskip-0.2cm\tx d^dx\,\sqrt{-h}\,\x^r(\car-2\L)=\frac{1}{\k^2}\int_{\pa\cm_{r_c}}\hskip-0.2cm\tx d^dx\,\sqrt{-h}\,\x^r\big(K^2-K^i_jK^j_i\big).
\ee
This redundancy in the way that the relevant boundary term can be parameterized is what fundamentally allows a meaningful comparison between the boundary counterterms and Kounterterms.

\subsubsection*{Universal boundary term restoring radial translations}

In order to render the variational problem well posed, it is necessary to formulate it in terms of conformal classes of boundary metrics, i.e. to restore radial diffeomorphisms as a symmetry of the theory on the regularized manifold $\cm_{r_c}$, at least asymptotically as $r_c\to\infty$. As we have argued, this can be achieved by projecting asymptotically onto AlAdS metrics and adding a suitable boundary term, $S\sbtx{ct}$. As we now review, the divergent part of this boundary term is universal: it is given by an asymptotic solution of the radial Hamilton-Jacobi equation \cite{Papadimitriou:2010as}. The only ambiguity in the boundary term amounts to the possibility of adding finite, local and covariant terms to $S\sbtx{ct}$, which is referred to as a choice of `renormalization scheme' in the context of the AdS/CFT correspondence. However, the divergent part of the boundary term that renders the variational problem well posed is completely unambiguous.  

The fact that, for any AlAdS metric, the canonical variables $h_{ij}$ and $K_{ij}$ are asymptotically related through a unique and universal relation allows us to take without loss of generality the boundary term $S\sbtx{ct}$ to be a function of the induced metric $h_{ij}$ and its transverse derivatives, i.e. derivatives with respect to the boundary coordinates $x^i$, but not of $\dot h_{ij}$. Since the radial cutoff does not break transverse diffeomorphisms $\x^{\perp i}(x)$ tangent to the cutoff surface, the boundary term should also preserve these. Namely, we demand that  
\be
\d_{\x^\perp}S\sbtx{ct}=-2\int_{\pa\cm_{r_c}}\hskip-0.2cm\tx d^dx\,\x_i^{\perp}D_j\Big(\frac{\d S\sbtx{ct}}{\d h_{ij}}\Big)=0,
\ee
which leads to the conservation equation   
\be\label{SB-div}
D_i\Big(\frac{\d S\sbtx{ct}}{\d h_{ij}}\Big)=0.
\ee
Finally, in order for the boundary term not to change the dynamics of the theory, we demand that it be local, i.e. polynomial in derivatives with respect to $x^i$. As we will see shortly, for even boundary dimension $d$, insisting on locality necessarily leads to a specific explicit dependence of $S\sbtx{ct}$ on the radial cutoff $r_c$, which is a manifestation of the holographic conformal anomaly \cite{Henningson:1998gx}.      

Writing the boundary term as    
\be
S\sbtx{ct}=\int_{\pa\cm_{r_c}}\hskip-0.2cm\tx d^dx\,\cl\sbtx{ct},
\ee
the sum of the regularized action and the boundary term transforms under diffeomorphisms as
\be\label{diff-var-Stot}
\d_\x \big(S_{\rm reg}+S\sbtx{ct}\big)=\int_{\pa\cm_{r_c}}\hskip-0.2cm\tx d^dx\,\x^r\mathscr{L}+\int_{\pa\cm_{r_c}}\hskip-0.2cm\tx d^dx\,\x^r(\dot\cl\sbtx{ct}-\pa_i\O\sbtx{ct}^i),
\ee
where $\mathscr{L}$ is given in \eqref{rLagrangian} and the vector density $\O\sbtx{ct}^i$ is implicitly determined by $\cl\sbtx{ct}$. Notice that, for compact $\pa\cm_{r_c}$, the density $\cl\sbtx{ct}$ is only defined up to a total derivative. However, $S\sbtx{ct}$ is unaffected by total derivative terms and so must be the variation \eqref{diff-var-Stot}. This determines that adding a total derivative term to $\cl\sbtx{ct}$, i.e. $\cl\sbtx{ct}\to\cl\sbtx{ct}+\pa_iv^i$, shifts $\O\sbtx{ct}^i$ according to $\O\sbtx{ct}^i\to \O\sbtx{ct}^i+\dot v^i$. The transformation \eqref{diff-var-Stot} implies that radial diffeomorphisms are restored as the radial cutoff is removed provided the r.h.s. vanishes, at least asymptotically, i.e. 
\be\label{ren-condition}
\lim_{r_c\to\infty}(\mathscr{L}+\dot\cl\sbtx{ct}-\pa_i\O\sbtx{ct}^i)=0.
\ee
This is the unintegrated version of the equivalent condition
\be\label{ren-condition-integrated}
\lim_{r_c\to\infty}(\dot S\sbtx{reg}+\dot S\sbtx{ct})=0,
\ee
which provides a universal expression for the boundary term  necessary to restore radial diffeomorphisms, as well as a systematic way for determining it. 

The key observation is that, on-shell, the regularized action, $S\sbtx{reg}$, is a covariant (but non local) functional of the induced metric, $h_{ij}$, on the radial cutoff, $\pa\cm_{r_c}$, and coincides with a specific solution, $\cs[h]$, of the radial Hamilton-Jacobi equation, which for pure AdS gravity takes the form  
\be\label{eq:HJ}
\frac{2\k^2}{\sqrt{-h}}\Big(h_{ik}h_{jl}-\frac{1}{d-1}h_{ij}h_{kl}\Big)\frac{\d\cs}{\d h_{kl}}\frac{\d\cs}{\d h_{ij}}+\frac{1}{2\k^2}\sqrt{-h}(\car-2\L)=0.
\ee
The condition \eqref{ren-condition-integrated}, therefore, implies that $S\sbtx{ct}$ takes the universal form   
\be\label{eq:universal-counterterm}
S\sbtx{ct}[h;r_c]=-\cs[h]+\text{finite as $r_c\to\infty$},
\ee
where $\cs[h]$ satisfies the Hamilton-Jacobi equation \eqref{eq:HJ}. It is a remarkable property of AlAdS manifolds that this quantity can be made local, i.e. polynomial in derivatives with respect to $x^i$, thus fulfilling also the locality requirement of $S\sbtx{ct}$, albeit at the expense of introducing explicit cutoff dependence in the case of even boundary dimension $d$.

\subsection{Boundary counterterms from the dilatation operator expansion} \label{dil_operator}

The result \eqref{eq:universal-counterterm} implies that the boundary term required to render the variational problem for AdS gravity well posed, and consequently the on-shell action finite, is given by the divergent part of a solution, $\cs[h]$, of the radial Hamilton-Jacobi equation. $S\sbtx{ct}$, therefore, can be determined by asymptotically solving the Hamilton-Jacobi equation \eqref{eq:HJ}. In simple cases, this can be done by enumerating all possible terms that can appear in $\cs[h]$, up to the desired order, and determining the coefficients using \eqref{eq:HJ} \cite{deBoer:1999xf,Martelli:2002sp,Elvang:2016tzz}. It is usually much more efficient, however, to solve the Hamilton-Jacobi equation systematically through the recursive relations obtained by a formal expansion of $\cs[h]$ in eigenfunctions of the dilation operator \cite{Papadimitriou:2004ap} (See \cite{Papadimitriou:2016yit} for a recent review and \cite{Papadimitriou:2011qb,Ross:2011gu,Chemissany:2014xpa} for generalizations to non conformal and non relativistic theories. A precursor of the dilatation operator method for pure AdS gravity was developed in \cite{Kraus:1999di}.) The original approach to holographic renormalization \cite{Henningson:1998gx,deHaro:2000xn} does not utilize the Hamilton-Jacobi equation and instead determines the asymptotic form of the regularized on-shell action by evaluating it explicitly on asymptotic solutions of the equations of motion. In the remaining of this section, we provide a brief, but self contained review of the dilatation operator method for solving the Hamilton-Jacobi equation in the case of pure AdS gravity. 

The Hamilton-Jacobi approach to gravity relies on the two Gauss-Codazzi equations
\be
\label{GC}
K^2-K^i_jK^j_i=\car-2\L,\qquad
D_iK^i_j-D_jK=0,
\ee
which correspond respectively to the $rr$ and $rj$ components of Einstein's equations. Upon using the relation \eqref{momentum} between the extrinsic curvature, $K_{ij}$, and the canonical momentum, $\p^{ij}$, these become respectively the Hamiltonian and momentum constraints  
\be\label{constraints}
\frac{2\k^2}{\sqrt{-h}}\left(\p^i_j\p^j_i-\frac{1}{d-1}\p^2\right)+\frac{1}{2\k^2}\sqrt{-h}\left(\car-2\L\right)=0,\qquad
D_j\p^{ij}=0.
\ee
The Hamilton-Jacobi equations for gravity are obtained from these constraints by writing the canonical momentum, $\p^{ij}$, as the gradient of a potential $\cs[h]$: 
\be\label{HJ-momentum}
\p^{ij}=\frac{\d\cs[h]}{\d h_{ij}}. 
\ee
In particular, the Hamiltonian constraint leads to the Hamilton-Jacobi equation \eqref{eq:HJ}, while the momentum constraint reflects the invariance of $\cs[h]$ under diffeomorphisms tangent to the constant $r$ surfaces. 

\subsubsection*{Dilatation operator}

When acting on covariant functionals of the induced metric, such as the Hamilton-Jacobi functional $\cs[h]$, the generator of radial translations may be represented as the functional operator
\be\label{radial_derivative}
\pa_r=\int \tx d^dx\,\dot{h}_{ij}[h]\frac{\d}{\d h_{ij}}.
\ee 
The dilatation operator is defined as the leading asymptotic form of the generator of radial translations in a covariant expansion as $r\to\infty$. Using the leading asymptotic behavior of the induced metric for AlAdS spacetimes in \eqref{FG-canonical}, we determine that $\dot{h}_{ij}\sim 2\ell^{-1}h_{ij}$ as $r\to\infty$. Hence, the leading asymptotic form of the generator of radial translations \eqref{radial_derivative} is given by
\be\label{dilatation-operator}
\pa_r\sim\ell^{-1}\int d^d x \;2 h_{ij} \frac{\d}{\d h_{ij}}\equiv\ell^{-1}\d\sbtx{D}.
\ee

\subsubsection*{Covariant expansion of the Hamilton-Jacobi functional}

The dilatation operator \eqref{dilatation-operator} enables us to expand the Hamilton-Jacobi functional $\cs[h]$ asymptotically, while maintaining manifest covariance. Writing  
\be\label{HJdensity}
\cs[h]=\int_{\pa\cm_{r_c}}\hskip-0.2cm\tx d^dx\;\cl[h],
\ee
we formally expand $\cl[h]$ in eigenfunctions of the dilatation operator as
\be\label{action_exp}
\cl=\cl\sub{0}+\cl\sub{2}+\cdots+\wt{\cl}\sub{d}\log{\rm e}^{-2r_c/\ell}+\cl\sub{d}+\cdots,
\ee
where
\be\label{dilatation-tranformations}
\d\sbtx{D}\cl\sub{2n}=(d-2n)\cl\sub{2n},\quad 0\leq n<d/2,\qquad \d\sbtx{D}\wt{\cl}\sub{d}=0.
\ee

The term $\wt{\cl}\sub{d}$ in the expansion \eqref{action_exp} is non zero only for even boundary dimension, $d$, and can be identified with the holographic conformal anomaly \cite{Henningson:1998gx}. The identification of the dilation operator with the leading asymptotic form of the generator of radial translations through \eqref{dilatation-operator} means that the relations \eqref{dilatation-tranformations} imply that $\cl\sub{2n} = \co(e^{(d-2n)r_c/\ell})$, $n<d/2$, and $\wt\cl\sub{d}=\co(1)$, as $r_c\to\infty$, and hence these terms are divergent as the cutoff is removed.\footnote{The relations \eqref{dilatation-tranformations}, however, contain more information. In particular, they require $\cl\sub{2n}$, $n<d/2$, and $\wt\cl\sub{d}$ to be homogeneous functionals of the induced metric $h_{ij}$.} Using \eqref{ren-condition-integrated}, therefore, we conclude that the boundary term that renders the variational problem well posed is given by 
\be\label{counterterms-v1}
S\sbtx{ct}[h;r_c]=-\int_{\pa\cm_{r_c}}\hskip-0.2cm\tx d^dx\big(\cl\sub{0}+\cl\sub{2}+\cdots+\wt{\cl}\sub{d}\log{\rm e}^{-2r_c/\ell}\big).
\ee

The term $\cl\sub{d}$ in the expansion \eqref{action_exp} has scaling dimension zero, i.e. $\cl\sub{d}=\co(1)$, as $r_c\to\infty$, and corresponds to the renormalized on-shell action. It is generically non local and cannot be determined from an asymptotic analysis alone. Moreover, it is not an eigenfunction of the dilatation operator in general. The action of $\d\sbtx{D}$ on $\cl\sub{d}$ can be deduced from the fact that $\cs[h]$, which is identified up to a constant with the regularized on-shell action, does not depend explicitly on the radial cutoff, due to the diffeomorphism invariance of the bulk action. Hence, the generator of cutoff translations, $\pa_{r_c}$, must act to leading order asymptotically as $\ell^{-1}\d\sbtx{D}$, namely
\be
\pa_{r_c}\big(\wt{\cl}\sub{d}\log e^{-2r_c/\ell}+\cl\sub{d}\big)\sim\ell^{-1} \d\sbtx{D}\big(\wt{\cl}\sub{d}\log e^{-2r_c/\ell}+\cl\sub{d}\big).
\ee
Using the identity $\d\sbtx{D}\sqrt{-h}=d\sqrt{-h}$, this implies that
\be\label{dilatation-tranformation-inhom}
\d\sbtx{D}\cl\sub{d}=-2\wt{\cl}\sub{d}.
\ee
However, $\cl\sub{d}$  does not play any role in the subsequent analysis of the present paper.

\subsubsection*{Recursion relations}

Our next task is to set up a recursive procedure for determining $\cl\sub{2n}$ for $n<d/2$ and $\wt\cl\sub{d}$. Given the expansion \eqref{action_exp} of $\cl$ in eigenfunctions of the dilatation operator, the canonical momentum \eqref{HJ-momentum} can be similarly expanded covariantly as  
\be
\label{momentum_exp}
\p^{ij}=\frac{\d}{\d h_{ij}}\int_{\pa\cm_{r_c}}\hskip-0.2cm\tx d^dx\;\cl
=\p_{(0)}^{ij}+\p_{(2)}^{ij}+\cdots+\wt{\p}_{(d)}^{ij}\log e^{-2r_c/\ell}+\p_{(d)}^{ij}+\cdots,
\ee
where
\be\label{momentum-coeffs}
\p^{ij}_{(2n)}=\frac{\d}{\d h_{ij}}\int_{\pa\cm_{r_c}}\hskip-0.2cm\tx d^dx\;\cl\sub{2n},\quad
\wt\p_{(d)}^{ij}=\frac{\d}{\d h_{ij}}\int_{\pa\cm_{r_c}}\hskip-0.2cm\tx d^dx\;\wt\cl\sub{d},\quad \p_{(d)}^{ij}=\frac{\d}{\d h_{ij}}\int_{\pa\cm_{r_c}}\hskip-0.2cm\tx d^dx\;\cl\sub{d}.
\ee 
The coefficients $\p_{(2n)}{}^i_j$, $n<d/2$, and $\wt\p_{(d)}{}^i_j$ are again eigenfunctions of the dilatation operator:
\bal
\d\sbtx{D}\p_{(2n)}{}^i_j=&\;(d-2n)\p_{(2n)}{}^i_j,  &&\d\sbtx{D}\p^{ij}_{(2n)}=(d-2n-2)\p^{ij}_{(2n)},\qquad n<d/2,\NO\\
\d\sbtx{D}\wt\p_{(d)}{}^i_j=&\;0, &&\hskip0.2cm\d\sbtx{D}\wt\p^{ij}_{(d)}=-2\wt\p^{ij}_{(d)}.
\eal

The key step in setting up a recursion procedure is to realize that the traces $\p\sub{2n}^i_i$ and $\wt\p\sub{d}^i_i$ are related algebraically with the coefficients $\cl\sub{2n}$ and $\wt\cl\sub{d}$ in the covariant expansion of $\cl$. The precise relation is obtained by applying $\d\sbtx{D}$ to $\cs[h]$. Using \eqref{HJdensity} and \eqref{HJ-momentum}, we obtain
\be 
\d\sbtx{D}\cs[h]=\int_{\pa\cm_{r_c}}\hskip-0.2cm\tx d^dx\,\d\sbtx{D}\cl=\int_{\pa\cm_{r_c}}\hskip-0.2cm\tx d^dx\,\d\sbtx{D} h_{ij}\p^{ij}=\int_{\pa\cm_{r_c}}\hskip-0.2cm\tx d^dx\,2\p^i_i,
\ee
where we have used the identity $\d\sbtx{D} h_{ij}=2h_{ij}$ in the last step. Since 
$\cl$ is only defined up to a total derivative, a suitable choice of the total derivative term allows us to write
\be\label{trace-id}
2\p^i_i=\d\sbtx{D}\cl.
\ee
Expanding both sides of this identity in eigenfunctions of the dilatation operator as in (\ref{action_exp}) and (\ref{momentum_exp}), we obtain (dropping summed indices in the traces)  
\bal\label{action-id}
&2\big(\p\sub{0}+\p\sub{2}+\cdots+\wt{\p}\sub{d}\log e^{-2r_c/\ell}+\p\sub{d}+\cdots\big)\NO\\
&=d\cl\sub{0}+(d-2)\cl\sub{2}+\cdots+0\cdot\wt{\cl}\sub{d}\log{\rm e}^{-2r_c/\ell}-2\wt{\cl}\sub{d}+0\cdot\cl\sub{d}+\cdots.
\eal
Equating terms of equal dilatation weight determines
\be\label{trace-relations} 
\cl\sub{2n}=\frac{2}{d-2n}\p\sub{2n},\quad 0\leq n<d/2,\qquad \wt\cl\sub{d}=-\p\sub{d},\qquad \wt \p\sub{d}=0. 
\ee

The relations \eqref{trace-relations} allow us to set up a recursion algorithm as follows. As we have seen, the leading asymptotic form of the induced metric in \eqref{FG-canonical} can be expressed in the form $\dot h_{ij}\sim 2\ell^{-1}h_{ij}$. Inserting this in the canonical momentum \eqref{momentum} gives
\be
\p\sub{0}^{ij}=\frac{(d-1)}{2\k^2\ell}\sqrt{-h}\;h^{ij},
\ee
which can be integrated to obtain
\be\label{L0}
\cl\sub{0}=\frac{(d-1)}{\k^2\ell}\sqrt{-h}.
\ee
Notice that these indeed satisfy the relations \eqref{trace-relations} for $n=0$.
Given these zeroth order expressions, the higher order terms can be computed iteratively by inserting the momentum expansion \eqref{momentum_exp} in the Hamiltonian constraint \eqref{constraints} and using the trace relations \eqref{trace-relations}. Matching terms of equal dilatation weight, we arrive at 
\be\boxed{
\label{linear-eq}
\cl\sub{2n}=\frac{\ell}{d-2n}\cq\sub{2n},\quad 0<n<d/2,\qquad \wt\cl\sub{d}=-\p\sub{d}=\left\{\begin{matrix} -\frac{\ell}{2} \cq\sub{d}, &  d\,\,\,\text{even,} \\ 0, & d\,\,\,\text{odd,}\end{matrix}\right. }
\ee
where
\bal\label{sources}
\cq\sub{2}&=\frac{\sqrt{-h}}{2\k^2}\car,\NO\\
\cq\sub{2n}&=\frac{2\k^2}{\sqrt{-h}}\sum_{m=1}^{n-1}\Big(\p\sub{2m}^i_j\p\sub{2n-2m}^j_i-\frac{1}{d-1}\p\sub{2m}\p\sub{2n-2m}
\Big), \qquad 1<n\leq d/2.
\eal
At order $n$, therefore, $\cq\sub{2n}$ and $\cl\sub{2n}$ are determined algebraically in terms of all $\p^{ij}_{(2m)}$ with $m<n$. From \eqref{momentum-coeffs}, we know that the functional derivative of $\cl\sub{2n}$ gives $\p^{ij}_{(2n)}$, which in turn allows us to obtain algebraically $\cq\sub{2n+2}$ and $\cl\sub{2n+2}$, thus completing the recursive procedure. This recursion algorithm systematically computes the boundary term \eqref{counterterms-v1} for any dimension $d$. 

The densities $\cq\sub{2n}$ and the symmetric tensor densities $\p^{ij}_{(2n)}$ are closely related to the study of conformal invariants in the mathematics literature, see e.g. \cite{Graham:1999jg,Anderson:2004yi,albin2005renormalizing,chang2005renormalized}. In particular, $\p^{ij}_{(2n)}$ is covariantly conserved for any $d$ and it is traceless for $d=2n$. The first of these properties follows immediately by covariantly expanding the momentum constraint in \eqref{constraints} in eigenfunctions of the dilatation operator, while the latter property is equivalent to the result $\wt\p\sub{d}=0$ in \eqref{trace-relations}. In combination with \eqref{momentum-coeffs}, this also implies that the integral of $\cq\sub{2n}$ over a compact $d=2n$ dimensional manifold is a conformal invariant. In the mathematics literature, $\cq\sub{d}$ is known as Branson's $Q$-curvature \cite{graham2001scattering,FG-Q-curvature,Q-curvature}.

The relation of $\cq\sub{2n}$ and $\p^{ij}_{(2n)}$ to conformal invariants explains why they are most compactly expressed in terms of curvature tensors that have a simple transformation under local Weyl  rescalings of $h_{ij}$, such as the Schouten tensor of $h_{ij}$ in $d$ dimensions (cf. the corresponding bulk tensors defined respectively in \eqref{BSchouten} and \eqref{BWeyl})
\be\label{bSchouten}
\cp_{ij}[h]=\frac{1}{d-2}\Big(\car_{ij}-\frac{1}{2(d-1)}\car h_{ij}\Big),
\ee
and the  Weyl tensor
\be\label{bWeyl}
\cw_{ikjl}[h]=\car_{ikjl}+h_{il}\cp_{kj}+h_{kj}\cp_{il}-h_{ij}\cp_{kl}-h_{kl}\cp_{ij},
\ee
which is traceless, i.e. $\cw^i{}_{kil}=0$, and transforms homogeneously under local Weyl transformations. Moreover, the Bianchi identity for the Riemann tensor
\be
D_p \car_{ijkl}+D_k\car_{ijlp}+D_l\car_{ijpk}=0,
\ee
implies that the Weyl tensor satisfies the Bianchi identity 
\be\label{bWeyl-Bianchi}
D_p \cw_{ijkl}+D_k\cw_{ijlp}+D_l\cw_{ijpk}+h_{ip}\cc_{jkl}+h_{ik}\cc_{jlp}+h_{il}\cc_{jpk}-h_{jp}\cc_{ikl}-h_{jk}\cc_{ilp}-h_{jl}\cc_{ipk}=0,
\ee
where $\cc_{ijk}$ is the Cotton tensor 
\be\label{bCotton}
\cc_{ijk}[h]=D_k\cp_{ij}-D_j\cp_{ik}.
\ee 

Recall that, in $d\geq 4$, a metric $h_{ij}$ is conformally flat if and only if $\cw_{ikjl}[h]=0$, while in $d=3$, the Weyl tensor is identically zero for all metrics and conformal flatness is instead equivalent to the vanishing of the Cotton tensor. All metrics in $d=2$ are conformally flat. Contracting appropriately the indices in \eqref{bWeyl-Bianchi} leads to the identity
\be\label{Weyl-Cotton-0}
D^i \cw_{ijkl}+(d-3)\cc_{jkl}+h_{jk}\cc^i{}_{il}-h_{jl}\cc^i{}_{ik}=0,
\ee
which in turn implies that
\be\label{Weyl-Cotton}
\frac{1}{d-3}D^k D^l\cw_{ikjl}=D^k \cc_{ijk}.
\ee

Implementing the recursion algorithm, we determine that the first few $\cl\sub{2n}$'s are given by
\bal\label{General-Ls}
\frac{2\k^2\ell}{\sqrt{-h}}\cl\sub{0}=&\;2(d-1),\NO\\
\frac{2\k^2\ell}{\sqrt{-h}}\cl\sub{2}=&\;\frac{\ell^2}{d-2}\car,\NO\\
\frac{2\k^2\ell}{\sqrt{-h}}\cl\sub{4}=&\;\frac{\ell^4}{(d-4)}\big(\cp^{ij}\cp_{ij}-\cp^2\big),\\
\frac{2\k^2\ell}{\sqrt{-h}}\cl\sub{6}=&\;\frac{\ell^6}{(d-6)(d-4)(d-2)}\Big[\cp_{ij}\cb^{ij}+(d-4)\Big(\cp^i_j\cp^j_k\cp^k_i-\cp\cp^i_j\cp^j_i-\frac{1}{2}\big(\cp^i_j\cp^j_i-\cp^2\big)\cp\Big)\Big],\NO
\eal
while the symmetric tensor densities $\p_{(2n)}^{ij}$, up to $n=2$, take the form
\bal\label{General-Pis}
\frac{2\k^2\ell}{\sqrt{-h}}\p^{ij}_{(0)}=&\;(d-1)h^{ij},\NO\\
\frac{2\k^2\ell}{\sqrt{-h}}\p^{ij}_{(2)}=&\;-\ell^2\big(\cp^{ij}-\cp h^{ij}\big),\NO\\
\frac{2\k^2\ell}{\sqrt{-h}}\p^{ij}_{(4)}=&\;-\frac{\ell^4}{(d-4)(d-2)}\Big[\cb^{ij}+(d-4)\Big(\cp^{ik}\cp_k^j-\cp\cp^{ij}-\frac12\big(\cp^{kl}\cp_{kl}-\cp^2\big)h^{ij}\Big)\Big].
\eal
The tensor $\cb^{ij}$ that appears in $\cl\sub{6}$ and $\p^{ij}_{(4)}$ is given by
\be\label{bBach}
\cb^{ij}=\Big(\frac{1}{d-3}D_k D_l+\cp_{kl}\Big)\cw^{ikjl}=D^k \cc_{ijk}+\cp_{kl}\cw^{ikjl},
\ee
and is known as the Bach tensor in dimension $d$ \cite{Bach,Graham:1999jg,ArlebackArticle,ArlebackThesis}. While $\cb^{ij}$ is traceless for any $d\geq 4$ and covariantly conserved for $d=4$, it is not covariantly conserved for $d>4$. However, the full expression for $\p^{ij}_{(4)}$ is conserved for any $d$ (but not traceless for $d>4$) and coincides with the modified Bach tensor introduced in \cite{Alaee:2018txy}. Higher order terms in the iterative procedure produce analogues of the Bach tensor for higher dimensions that are more than quadratic in the curvatures.

\subsubsection*{$\mathbf{Q}$-curvature and conformal anomaly}

The covariant densities $\cq\sub{2n}$ determined by the recursion relations
\eqref{linear-eq}-\eqref{sources} are functions of the boundary dimension, $d$. When $d$ is even, $\cq\sub{d}$ corresponds to the holographic conformal anomaly  \cite{Henningson:1998gx}, as well as Branson's $Q$-curvature \cite{Q-curvature,graham2001scattering,FG-Q-curvature}. A conjecture by Deser and Schwimmer \cite{Deser:1993yx} for the general structure of conformal anomalies, later proven by Alexakis \cite{alexakis2007decomposition} specifically for the $Q$-curvature, implies that $\wt\cl\sub{d}$ can be  decomposed as 
\be\label{Q-curvature}
\wt\cl\sub{d}= a\,\ce_d[h]+\sum_{I=1}^{N_d} c_I\, \ci_I[h]+\pa_i(\sqrt{-h}\,\cj^i[h]),
\ee
where $\ce_d[h]$ is the Euler density in $d$ dimensions, $\ci_I[h]$, $I=1,\cdots, N_d$, are all {\em local} conformal invariants in $d$ dimensions, and $\cj^i[h]$ is a globally defined (and renormalization scheme dependent) total derivative term that does not contribute to the integral of $\wt\cl\sub{d}$ over a compact boundary. While in the conjecture of Deser and Schwimmer for the conformal anomaly the coefficients $a$, $c_I$ are generic and depend on the specific conformal field theory, in the $Q$-curvature their values are related and are all proportional to the inverse gravitational constant $\k^{-2}\propto G^{-1}$. 

While the Euler density is the unique global conformal invariant in every even dimension (type-A anomaly in the classification of \cite{Deser:1993yx}), there can be multiple local conformal invariants (type-B anomaly) in any even dimension (see e.g. \cite{Boulanger:2018rxo} and references therein for the general classification). There exist no local conformal invariants in two dimensions and so, in that case, the $Q$-curvature is proportional to the Euler density, namely
\be
\wt\cl\sub{2}=-\frac{\p\ell}{\k^2}\ce_2,
\ee 
where $\ce_{2n}$ is the Pfaffian of the Riemann curvature of the induced metric $h_{ij}$ as normalized in \eqref{Pf-index}. In $d=4$, the unique local conformal invariant is the square of the Weyl tensor $\cw_{ijkl}\cw^{ijkl}$. From the expression for $\wt\cl\sub{4}$ obtained above, one finds that it can be expressed in the form \eqref{Q-curvature} as
\be
\wt\cl\sub{4}=\frac{2\ell^3}{\k^2}\Big(\frac{\p^2}{2}\ce_4-\frac{1}{64}\sqrt{-h}\,\cw^{ijkl}\cw_{ijkl}\Big).
\ee

There are three independent local conformal invariants in six dimensions. A suitable basis is \cite{fefferman1985mathematical,parker1987,Erdmenger:1997gy,Graham:1999jg}\footnote{As it was pointed out in \cite{Bastianelli:2000hi,Tseytlin:2000sf}, the expression for $\ci_3$ in \cite{Bonora:1985cq,Deser:1993yx} that is often quoted in the physics literature is incorrect and  does not transform homogeneously under local Weyl transformations. We thank Kostas Skenderis for communication on this point.}
\bal\label{d=6-Weyl-invariants}
\ci_1=&\;\sqrt{-h}\,\cw^{ij}{}_{kl}\cw^{kl}{}_{pq}\cw^{pq}{}_{ij},\NO\\
\ci_2=&\;\sqrt{-h}\,\cw^{ij}{}_{kl}\cw^{kp}{}_{iq}\cw^{lq}{}_{jp},\NO\\
\ci_3=&\;\sqrt{-h}\big(\cv_{ijklm}\cv^{ijklm}-16\cw^{ijkl}D_i\cc_{jkl}+16\cw^{ijkl}\cp_{im}\cw^m{}_{jkl}+16\cc_{ijk}\cc^{ijk}\big),
\eal
where
\be
\cv_{ijklm}=D_m\cw_{ijkl}+h_{im}\cc_{jkl}-h_{jm}\cc_{ikl}+h_{km}\cc_{lij}-h_{lm}\cc_{kij}.
\ee
One can check that, in this case, the $Q$-curvature can be written in the form
\be
\wt\cl\sub{6}=\frac{3\ell^5}{2\k^2}\Big(-\frac{\p^3}{6}\ce_6+\frac{1}{2304}(7\ci_1+4\ci_2-3\ci_3)\Big)+\text{total derivative}.
\ee

We conclude this subsection with an observation that will be important for the comparison with the Kounterterms later on. Notice that, for conformally flat boundary metrics, $h_{ij}$, all local conformal invariants vanish identically and so the only non trivial contribution to the $Q$-curvature in that case is from the Euler density.

\subsection{Renormalization scheme dependence} 
\label{scheme}

The above analysis shows that all covariant counterterms that cancel divergences are both unique and universal for a given bulk action. However, so far we have not discussed possible finite counterterms. As we now briefly review, there is an ambiguity in the choice of finite counterterms that corresponds to the renormalization scheme dependence of the holographic dual field theory.     

The punchline of the above analysis is that the boundary counterterms take the form  
\be\label{counterterms}\boxed{ 
S\sbtx{ct}[h;r_c]=\int_{\pa\cm_{r_c}}\hskip-0.2cm\tx d^dx\big(\cl_{\rm ct}\sub{0}+\cl_{\rm ct}\sub{2}+\cdots+\wt{\cl}_{\rm ct}\sub{d}\log{\rm e}^{-2r_c/\ell}+\cl_{\rm ct}\sub{d}\big),}
\ee
where 
\be
\cl_{\rm ct}\sub{0}=-\cl\sub{0},\quad\ldots,\quad \wt{\cl}_{\rm ct}\sub{d}=-\wt{\cl}\sub{d},
\ee
and $\cl\sub{0}$, $\cl\sub{2}$, \ldots, $\wt\cl\sub{d}$ are uniquely determined by the recursion relations \eqref{linear-eq}. Crucially, these recursion relations do not determine $\cl\sub{d}$, which is non local and corresponds to the renormalized on-shell action. The finite term $\cl_{\rm ct}\sub{d}$ in the counterterms \eqref{counterterms} is {\em not} related to $\cl\sub{d}$. Instead, $\cl_{\rm ct}\sub{d}$ may be set to zero, or it can be any local and covariant density whose integral over the boundary is a conformal invariant. In other words, $\cl_{\rm ct}\sub{d}$ is in general a global conformal invariant. Terms that do not preserve Weyl invariance or even covariance with respect to boundary diffeomorphisms may also be added, but such terms would introduce cohomologically trivial contributions to the conformal and gravitational anomalies of the dual field theory and hence they should not be included in $\cl_{\rm ct}\sub{d}$.

For pure gravity, $\cl_{\rm ct}\sub{d}$ can be non-zero only for even boundary dimension $d$, since for odd $d$ there exist no local densities that satisfy the above criteria. The structure of $\cl_{\rm ct}\sub{d}$ for even $d$ is analogous to that of the $Q$-curvature \eqref{Q-curvature}, namely
\be\label{finite-ct}
\cl_{\rm ct}\sub{d}= s_0\,\ce_d[h]+\sum_{I=1}^{N_d} s_I\, \ci_I[h].
\ee
However, the constants $s_0$, $s_I$ here can be chosen at will, while adding a total derivative term does not have any consequence. The only ambiguity in the boundary counterterms is the choice of these constants, which corresponds to the renormalization scheme dependence of the dual field theory.

\section{Kounterterms in AdS gravity}
\label{sec:Kounterterms}

The algorithm described in the previous section recursively determines the boundary counterterms in any dimension. However, the complexity of the counterterms for generic AlAdS manifolds increases rapidly with the dimension and there exists no closed form expression valid for arbitrary dimension. The boundary Kounterterms are an attempt to remedy this. First proposed for even bulk dimensions ($d$ odd) in \cite{Aros:1999kt,Olea:2005gb} and later generalized to odd bulk dimensions in \cite{Olea:2006vd}, the Kounterterms constitute a closed form expression for a boundary term applicable to any dimension. As we will review in this section, they are intimately related to topological aspects of conformally compact Einstein manifolds, which were independently studied in the mathematics literature at around the same time \cite{Anderson_l2,albin2005renormalizing,chang2005renormalized}.     

The Kounterterms correspond to adding to the action \eqref{action} a boundary term of the form  
\be\label{Kounterterms}
S\sbtx{K}=\int_{\pa \cm_\e}\tx d^dx\,\cl\sbtx{K}=c_d\int_{\partial
\cm_\e}\tx d^dx\,B_{d}[h,K,\mathcal{R}]-S\sbtx{GH} \,,
\ee%
where $c_{d}$ is a dimension dependent constant and $B_{d}[h,K,\mathcal{R}]$ is a density polynomial in the extrinsic and intrinsic curvatures of the regularized boundary $\pa\cm_\e$ that takes different form for even and odd dimensions. As we will see, the reason why $c_d$ is kept explicit is so that the normalization of $B_d$ matches certain bulk topological invariants. Notice that the negative of the Gibbons-Hawking term in the boundary term \eqref{Kounterterms} is designed to cancel the Gibbons-Hawking term in \eqref{action}. 

Contrary to the counterterms \eqref{counterterms}, the Kounterterms depend explicitly on both the induced metric, $h_{ij}$, and the extrinsic curvature, $K_{ij}$, or equivalently the canonical momentum, $\p^{ij}$. However, we saw in section \ref{sec:hr} that the variational problem on AlAdS spaces must be formulated within the space of asymptotic solutions of the equations of motion, and so $K_{ij}$ and $h_{ij}$ are asymptotically related. By inserting the asymptotic relation $K_{ij}[h]$ (obtained by solving the bulk equations of motion or the radial Hamilton-Jacobi equation) in the Kounterterms \eqref{Kounterterms}, one obtains a boundary term that is a function of the metric $h_{ij}$ only and can be compared directly with the counterterms \eqref{counterterms}. Since the Kounterterms are polynomial in the extrinsic and intrinsic curvatures, their divergent part is guaranteed to be a local and covariant expression when the asymptotic relation $K_{ij}[h]$ is incorporated. A priori, the finite part in the covariant expansion of the Kounterterms could still be  non-local, but we will show in this section that this is not the case.    

\subsection{Chern-Gauss-Bonnet theorem for manifolds with boundary}

In order to appreciate the origin and nature of the boundary Kounterterms, it is useful to briefly recall the generalized Chern-Gauss-Bonnet theorem for manifolds with boundary \cite{Chern1945}. Given a $d+1$ dimensional manifold, $\cm_\e$, we introduce the $d$-forms\footnote{The forms $\F_k$ here differ from those defined in \cite{Chern1945} by a factor of $(-1)^{d+k}$, due to a different sign in the definition of the connection one-form and curvature two-form.}
\be\label{Chern-Phi}
\F_k=\e_{a_1a_2\cdots a_{d}d+1}\Hat R^{a_1a_2}\wedge\Hat R^{a_3a_4}\wedge\cdots\wedge\Hat R^{a_{2k-1}a_{2k}}\wedge\Hat\o^{a_{2k+1}d+1}\wedge\Hat\o^{a_{2k+2}d+1}\wedge\cdots\wedge\Hat \o^{a_{d}d+1},
\ee
where $a_1$, $a_2$,$\cdots$ are tangent space indices, $\e_{a_1a_2\cdots a_{d}a_{d+1}}$ is the Levi-Civita tensor, $\Hat R^{ab}$ and $\Hat\o^{ab}$ are respectively the curvature two-form and connection one-form, and the integer $k$ takes the values $k=0,1,\cdots,\left[\frac{d+1}{2}\right]-1$, with $[x]$ indicating the integer part of $x$.

Chern showed that the $d$-form 
\be\label{CS-form}
\P=-\frac{1}{2^{d+1}\p^{\frac d2}}\sum_{k=0}^{[\frac d2]}\frac{1}{k!\,\G\big(\frac{d}{2}+1-k\big)}\F_k,
\ee
satisfies 
\be
-\tx d\P=\O,
\ee
where the $(d+1)$-form $\O$ is the Euler-Poincar\'e density when $d+1$ is even and zero otherwise:
\be\label{Pf-Chern}
\O=\left\{\begin{array}{ll} \text{Pf}(\Hat R), & \text{if $d+1=2n$ is even},\\
0, & \text{if $d+1$ is odd},\end{array}\right.
\ee
with the Pfaffian of the curvature two-form $\text{Pf}(\Hat R)$ given by\footnote{As for $\F_k$ defined above, this agrees with \cite{Chern1945} once the different sign in the definition of the curvature two-form is taken into account.} 
\be\label{Pf}
\text{Pf}(\Hat R)\equiv\frac{1}{(4\p)^n n!}\e_{a_1a_2\cdots a_{d+1}}\Hat R^{a_1a_2}\wedge\Hat R^{a_3a_4}\wedge\cdots\wedge\Hat R^{a_{d}a_{d+1}},\qquad d+1=2n.
\ee
The Chern-Gauss-Bonnet theorem states that the Euler-Poincar\'e characteristic for a manifold $\cm_\e$ with boundary $\pa\cm_\e$ is given by
\be\label{CGB-theorem}
\c(\cm_\e)=\int_{\cm_\e}\O+\int_{\pa\cm_\e}\P.
\ee

We note for later reference that the integral of the Pfaffian over an even dimensional manifold $\cm_\e$ can be written in coordinate basis as
\be\label{Pf-index}
\int_{\cm_\e} \text{Pf}(\Hat R)=\frac{1}{2^n(4\p)^n n!}\int_{\cm_\e}\tx d^{2n}x\,\d_{\m_1\cdots\m_{2n}}^{\n_1\cdots\n_{2n}}\sqrt{-g}\,R^{\m_1\m_2}{}_{\n_1\n_1}\cdots R^{\m_{2n-1}\m_{2n}}{}_{\n_{2n-1}\n_{2n}},
\ee
where
\be\label{antisym-deltas}
\d_{\m_1\cdots\m_{2n}}^{\n_1\cdots\n_{2n}}\equiv (2n)!\,\d_{\m_1}^{[\n_1}\cdots\d^{\n_{2n}]}_{\m_{2n}}=\sum_{P\in S_{2n}}{\rm sgn}(P)\d^{\n_1}_{P(\m_1)}\cdots\d^{\n_{2n}}_{P(\m_{2n})},
\ee 
is the totally antisymmetrized product of Kronecker deltas. Moreover, choosing the orientation of $\pa\cm_\e$ in $\cm_\e$ so that $\lbar\ve_{i_{1}\cdots i_d}\equiv\ve_{r i_{1}\cdots i_d}$, the pullback of the $d$-forms \eqref{Chern-Phi} on  $\pa\cm_\e$ is given by
\bal\label{Chern-Phi-pullback}
\left.\F_k\right|_{\pa\cm_\e}=&\;(-1)^d\sqrt{-h}\,\d^{j_1\cdots j_d}_{i_1\cdots i_d}\Big(\frac12\car^{i_1i_2}{}_{j_1j_2}-K^{i_1}_{j_1}K^{i_2}_{j_2}\Big)\times\NO\\
&\;\cdots\times\Big(\frac12\car^{i_{2k-1}i_{2k}}{}_{j_{2k-1}j_{2k}}-K^{i_{2k-1}}_{j_{2k-1}}K^{i_{2k}}_{j_{2k}}\Big)\times K^{i_{2k+1}}_{j_{2k+1}}\cdots K^{i_d}_{j_d}\,. 
\eal
It follows that the pullback of $\P$ on $\pa\cm_\e$ takes the form
\bal\label{pullbackPi}
\left.\P\right|_{\pa\cm_\e}=&\;\frac{(-1)^{d+1}}{2^{d+1}\p^{\frac d2}}\sqrt{-h}\,\d^{j_1\cdots j_d}_{i_1\cdots i_d}\sum_{k=0}^{[\frac d2]}\frac{1}{k!\,\G\big(\frac{d}{2}+1-k\big)}\Big(\frac12\car^{i_1i_2}{}_{j_1j_2}-K^{i_1}_{j_1}K^{i_2}_{j_2}\Big)\times\NO\\
&\;\cdots\times\Big(\frac12\car^{i_{2k-1}i_{2k}}{}_{j_{2k-1}j_{2k}}-K^{i_{2k-1}}_{j_{2k-1}}K^{i_{2k}}_{j_{2k}}\Big)\times K^{i_{2k+1}}_{j_{2k+1}}\cdots K^{i_d}_{j_d}\,.
\eal
As we review in the next subsection, the Kounterterms for even dimensional AlAdS manifolds are directly related to this expression.

\subsection{Kounterterms and their topological origin}

We are now ready to discuss the boundary Kounterterms and their relation with the Chern form \eqref{CS-form}. Since their defining expressions are different for even and odd dimensions, we consider these two cases separately. We will see in section \ref{sec:conflat}, however, that for conformally flat manifolds, the Kounterterms for even and odd dimensions coincide, up to finite local terms.   

\subsubsection*{Even dimensions} 

For even dimensional AlAdS manifolds ($d$ odd), the density $B_{2n-1}$ defining the Kounterterms \eqref{Kounterterms} is given by \cite{Olea:2005gb}
\bal\label{Beven}
B_{2n-1}[h,K,\mathcal{R}] =&\; 2n\sqrt{-h }\int_{0}^{1}\tx dt\,\delta_{j_{1\ldots }j_{2n-1}}^{i_{1}\ldots
i_{2n-1}}K_{i_{1}}^{j_{1}}\Big(\frac{1}{2}\,\mathcal{R}^{j_{2}j_{3}}{}
_{i_{2}i_{3}}
-t^{2}K_{i_{2}}^{j_{2}}K_{i_{3}}^{j_{3}}\Big) \times \notag \\
&\cdots \times \Big( \frac{1}{2}\,\mathcal{R}^{j_{2n-2}j_{2n-1}}{}
_{i_{2n-2}i_{2n-1}}
-t^{2}K_{i_{2n-2}}^{j_{2n-2}}K_{i_{2n-1}}^{j_{2n-1}}\Big) \,,
\eal
while the value of the proportionality constant $c_{2n-1}$ is
\be
\label{c-even}
c_{2n-1}=\frac{(-\ell
^2)^{n-1}}{2\k^2n(2n-2)!}\,.
\ee
The expression \eqref{Beven} contains a parametric integral which comes from the use of the Cartan homotopy operator in order to find the boundary term which is locally equivalent to the Euler term in the bulk. Thus, $B_{2n-1}$ is a Chern-Simons-like density associated to the Lorentz group, which naturally incorporates a second spin connection in order to restore covariance at the boundary \cite{Eguchi:1980jx}.

The origin of the density \eqref{Beven} becomes more transparent by the observation that it is proportional to the pullback of the Chern form $\P$ on $\pa\cm_\e$ given in \eqref{pullbackPi}, namely 
\be\label{Beven-CS}
B_{2n-1}=(4\pi)^n n!\;\left.\P\right|_{\pa\cm_\e}.
\ee
Hence, in even dimensions, the Chern-Gauss-Bonnet theorem \eqref{CGB-theorem} takes the form  
\be
\chi(\cm_\e)= \int_{\cm_\e} \text{Pf}(\Hat R)+ \frac{1}{(4\pi)^n n!}\int_{\pa \cm_\e}\tx d^{2n-1}x\,B_{2n-1}.
\ee

The identity \eqref{Beven-CS} can be easily proven by rearranging the expressions \eqref{pullbackPi} and \eqref{Beven}. Using the binomial expansion and performing the integral over the parameter $t$ in \eqref{Beven} leads to 
\bal\label{Beven-expanded}
B_{2n-1} =&\; \sqrt{-h }\,\delta_{j_{1\ldots }j_{2n-1}}^{i_{1}\ldots
i_{2n-1}}\sum_{k=0}^{n-1}\frac{(-1)^{k}n!}{2^{n-k-2}(2k+1)k!(n-k-1)!} K_{i_{1}}^{j_{1}}K_{i_{2}}^{j_{2}}K_{i_{3}}^{j_{3}}\cdots K_{i_{2k}}^{j_{2k}}K_{i_{2k+1}}^{j_{2k+1}}\NO\\
&\;\times \mathcal{R}^{j_{2k+2}j_{2k+3}}{}_{i_{2k+2}i_{2k+3}}\cdots\mathcal{R}^{j_{2n-2}j_{2n-1}}{}_{i_{2n-2}i_{2n-1}}.
\eal
Similarly, applying the binomial expansion and specializing \eqref{pullbackPi} to $d=2n-1$, we obtain
\bal\label{pullbackPi-even-expanded}
\left.\P\right|_{\pa\cm_\e}=&\;\frac{(-1)^{d+1}}{2^{d+1}\p^{\frac d2}}\sqrt{-h}\,\d^{j_1\cdots j_d}_{i_1\cdots i_d}\sum_{m=0}^{[\frac d2]}\sum_{k=m}^{[\frac d2]}\frac{(-1)^{k-m}}{2^m(k-m)!m!\,\G\big(\frac{d}{2}+1-k\big)}\car^{i_1i_2}{}_{j_1j_2}\cdots\car^{i_{2m-1}i_{2m}}{}_{j_{2m-1}j_{2m}}\times\NO\\
&\;\times K^{i_{2m+1}}_{j_{2m+1}}K^{i_{2m+2}}_{j_{2m+2}}\cdots  K^{i_d}_{j_d}\NO\\
=&\;\frac{1}{(4\p)^{n}}\sqrt{-h}\,\d^{j_1\cdots j_{2n-1}}_{i_1\cdots i_{2n-1}}\sum_{m=0}^{n-1}\frac{(-1)^{n-1-m}}{2^{m-1} (2n-1-2m)(n-1-m)!m!}\car^{i_1i_2}{}_{j_1j_2}\cdots\car^{i_{2m-1}i_{2m}}{}_{j_{2m-1}j_{2m}}\times\NO\\
&\;\times K^{i_{2m+1}}_{j_{2m+1}}K^{i_{2m+2}}_{j_{2m+2}}\cdots  K^{i_{2n-1}}_{j_{2n-1}}\NO\\
=&\;\frac{1}{(4\p)^n}\sqrt{-h}\,\d^{j_1\cdots j_{2n-1}}_{i_1\cdots i_{2n-1}}\sum_{k=0}^{n-1}\frac{(-1)^{k}}{2^{n-k-2} (2k+1)(n-1-k)!k!}K_{i_{1}}^{j_{1}}K_{i_{2}}^{j_{2}}K_{i_{3}}^{j_{3}}\cdots K_{i_{2k}}^{j_{2k}}K_{i_{2k+1}}^{j_{2k+1}}\NO\\
&\;\times \mathcal{R}^{j_{2k+2}j_{2k+3}}{}_{i_{2k+2}i_{2k+3}}\cdots\mathcal{R}^{j_{2n-2}j_{2n-1}}{}_{i_{2n-2}i_{2n-1}}\,.
\eal
Comparing this expression with \eqref{Beven-expanded} results in \eqref{Beven-CS}.

\subsubsection*{Odd dimensions}

The Chern form \eqref{CS-form} is defined for both even and odd dimensions. As in even dimensions, therefore, a natural candidate for the Kounterterms in odd bulk dimensions is the pullback \eqref{pullbackPi} of $\P$ on $\pa\cm_\e$. However, specializing \eqref{pullbackPi} to $d=2n$ gives
\bal\label{pullbackPi-odd-expanded}
\left.\P\right|_{\pa\cm_\e}=&\;-\frac{1}{2^{2n+1}\p^{n}}\sqrt{-h}\,\d^{j_1\cdots j_{2n}}_{i_1\cdots i_{2n}}\sum_{m=0}^{n}\sum_{k=0}^{n-m}\frac{(-1)^{k}}{2^m m!k!(n-m-k)!}\car^{i_1i_2}{}_{j_1j_2}\cdots\car^{i_{2m-1}i_{2m}}{}_{j_{2m-1}j_{2m}}\times\NO\\
&\;\times K^{i_{2m+1}}_{j_{2m+1}}K^{i_{2m+2}}_{j_{2m+2}}\cdots  K^{i_{2n}}_{j_{2n}}\NO\\
=&\;-\frac12\frac{1}{2^n(4\p)^nn!}\sqrt{-h}\,\d^{j_1\cdots j_{2n}}_{i_1\cdots i_{2n}}\car^{i_1i_2}{}_{j_1j_2}\cdots\car^{i_{2n-1}i_{2n}}{}_{j_{2n-1}j_{2n}}=-\frac12\ce(\car),
\eal
where $\ce(\car)$ is the Euler density of $\pa\cm_\e$, which is independent of the regulator $\e$ and hence finite. In particular, the Chern-Gauss-Bonnet theorem \eqref{CGB-theorem} stipulates that the Euler characteristic of an odd dimensional manifold with boundary is given by the Euler characteristic of the boundary. It follows that $\P$ cannot provide a suitable boundary term for odd bulk dimensions. 

The Kounterterms for odd bulk dimensions proposed in \cite{Olea:2006vd} instead take the form
\bal\label{Bodd}
B_{2n}[h,K,\mathcal{R}]=&\;2n\sqrt{-h }\int_{0}^{1}\tx dt\int_{0}^{t}\tx ds\,\d_{j_{1}\ldots j_{2n}}^{i_{1}\ldots i_{2n}}K_{i_{1}}^{j_{1}}\d_{i_{2}}^{j_{2}}\Big(\frac{1}{2}\,\mathcal{R}^{j_{3}j_{4}}{}_{i_{3}i_{4}}-t^{2}K_{i_{3}}^{j_{3}}K_{i_{4}}^{j_{4}}+\frac{%
s^{2}}{\ell ^{2}}\,\d_{i_{3}}^{j_{3}}\d_{i_{4}}^{j_{4}}\Big)
\times \notag \\
&\;\cdots \times \Big( \frac{1}{2}\,\mathcal{R}^{j_{2n-1}j_{2n}}{}
_{i_{2n-1}i_{2n}}-t^{2}K_{i_{2n-1}}^{j_{2n-1}}K_{i_{2n}}^{j_{2n}}+%
\frac{s^{2}}{\ell ^{2}}\,\d_{i_{2n-1}}^{j_{2n-1}}\d_{i_{2n}}^{j_{2n}}\Big), 
\eal
with proportionality constant
\be
\label{c-odd}
c_{2n}=\frac{( -\ell ^{2}) ^{n-1}}{%
2^{2n-1}\k^2n(n-1)!^{2}}\,.
\ee
A key difference between \eqref{Beven} and \eqref{Bodd} is that the former does not explicitly depend on the AdS radius $\ell$, which is a direct consequence of its topological origin. In contrast, \eqref{Bodd} is not related to a topological quantity and differs from the pullback \eqref{pullbackPi} of the Chern form on $\pa\cm_\e$. Using the binomial expansion and integrating over the parameters $s$ and $t$ in \eqref{Bodd}, we obtain
\bal\label{Bodd-expanded}
B_{2n}=&\;\sqrt{-h }\,\d_{j_{1}\ldots j_{2n}}^{i_{1}\ldots i_{2n}}K_{i_{1}}^{j_{1}}\d_{i_{2}}^{j_{2}}\sum_{m=0}^{n-1}\sum_{k=0}^m\frac{n!(-1)^{m-k}\ell^{-2n+2+2m}}{2^{k}(n-k)(2n-1-2m)k!(m-k)!(n-1-m)!} \times\\
&\;\mathcal{R}^{j_{3}j_{4}}{}_{i_{3}i_{4}}\cdots\mathcal{R}^{j_{2k+1}j_{2k+2}}{}
_{i_{2k+1}i_{2k+2}}\,K_{i_{2k+3}}^{j_{2k+3}}K_{i_{2k+4}}^{j_{2k+4}}
\cdots K_{i_{2m+1}}^{j_{2m+1}}K_{i_{2m+2}}^{j_{2m+2}}\, 
\d_{i_{2m+3}}^{j_{2m+3}}\d_{i_{2m+4}}^{j_{2m+4}}\cdots\d_{i_{2n-1}}^{j_{2n-1}}\d_{i_{2n}}^{j_{2n}}\,. \NO
\eal
Clearly, this expression is different from the pullback of $\P$ on $\pa\cm_\e$ in \eqref{pullbackPi-odd-expanded}.

\subsection{Kounterterms in terms of intrinsic boundary curvature}
 
Since the Kounterterms depend explicitly on the extrinsic curvature, $K^i_j$, they cannot be compared directly with the boundary counterterms \eqref{counterterms}, which only depend on the intrinsic curvature of the induced metric $h_{ij}$. 
However, in section \ref{sec:hr} we argued that the variational problem on AlAdS spaces must be formulated within the space of asymptotic solutions of the equations of motion, which implies that $K_{ij}$ and $h_{ij}$ are asymptotically related. Using this relation, the Kounterterms \eqref{Kounterterms} become a function of the metric $h_{ij}$ only and can be compared with the counterterms \eqref{counterterms}. 

The asymptotic on-shell relation $K_{ij}[h]$ between the extrinsic curvature and the induced metric follows from the asymptotic solution of the Hamilton-Jacobi equation that determines the boundary counterterms. In particular, the definition of the canonical momentum $\p^{ij}$ in \eqref{momentum} and its covariant expansion \eqref{momentum_exp} imply that the extrinsic curvature too can be expanded in eigenfunctions of the dilatation operator as \cite{Papadimitriou:2004ap}
\be\label{K-exp-generic}
K^i_j=K\sub{0}^i_j+K\sub{2}^i_j+\cdots+\wt K\sub{d}^i_j\log e^{-2r_c/\ell}+K\sub{d}^i_j+\cdots,
\ee     
where each term is related with the corresponding one in \eqref{momentum_exp} through the identities
\be\label{K-momentum-coeffs}
K^{ij}_{(2n)}=-\frac{2\k^2}{\sqrt{-h}}\Big(\p_{(2n)}^{ij}-\frac{1}{d-1}h^{ij}\p_{(2n)}\Big),\quad n\leq d,\qquad \wt K^{ij}_{(d)}=-\frac{2\k^2}{\sqrt{-h}}\Big(\wt\p_{(d)}^{ij}-\frac{1}{d-1}h^{ij}\wt\p_{(d)}\Big).
\ee
Inserting the first few orders of the canonical momentum coefficients given in \eqref{General-Pis}, one obtains the covariant expansion of the extrinsic curvature in \eqref{K-exp}.  
 
The Kounterterms \eqref{Kounterterms} can be expressed in terms of the density
\be\label{Kounterterms-density}
\cl\sbtx{K}=c_d B_{d}[h,K,\mathcal{R}]-\frac{1}{\k^2}\sqrt{-h}\,K\,,
\ee%
where $B_{d}$ and $c_d$ are given in \eqref{Beven} and \eqref{c-even} for even boundary dimension $d$ and in \eqref{Bodd} and \eqref{c-odd} for odd $d$. Integrating over the auxiliary parameters $t$ and $s$ in the definition of $B_d$ (or equivalently using \eqref{Beven-expanded} and \eqref{Bodd-expanded}), we determine that, up to dimension six, $B_d$ takes the form 
\bal
B_{2}=&\;\sqrt{-h }\,\d_{j_{1}j_{2}}^{i_{1}i_{2}}K_{i_{1}}^{j_{1}}\d_{i_{2}}^{j_{2}}, \NO\\
B_{3}=&\; \sqrt{-h }\,\delta_{j_{1\ldots }j_{3}}^{i_{1}\ldots
i_{3}}K_{i_{1}}^{j_{1}}\Big(2\mathcal{R}^{j_{2}j_{3}}{}_{i_{2}i_{3}}-\frac{4}{3} K_{i_{2}}^{j_{2}}K_{i_{3}}^{j_{3}}\Big),\NO\\
B_{4}=&\;\sqrt{-h }\,\d_{j_{1}\ldots j_{4}}^{i_{1}\ldots i_{4}}K_{i_{1}}^{j_{1}}\d_{i_{2}}^{j_{2}}\Big(\mathcal{R}^{j_{3}j_{4}}{}
_{i_{3}i_{4}}-K_{i_{3}}^{j_{3}}K_{i_{4}}^{j_{4}}+\frac{%
1}{3\ell ^{2}}\,\d_{i_{3}}^{j_{3}}\d_{i_{4}}^{j_{4}}\Big),\NO\\
B_5=&\;\sqrt{-h }\,\d_{j_{1}\ldots j_{5}}^{i_{1}\ldots i_{5}}K_{i_{1}}^{j_{1}}\Big(\frac32\car^{j_2j_3}{}_{i_2i_3}\car^{j_4j_5}{}_{i_4i_5}-2\car^{j_2j_3}{}_{i_2i_3}K^{j_4}_{i_4}K^{j_5}_{i_5}+\frac65K^{j_2}_{i_2}K^{j_3}_{i_3}K^{j_4}_{i_4}K^{j_5}_{i_5}\Big),\NO\\
B_{6}=&\;\sqrt{-h }\,\d_{j_{1}\ldots j_{6}}^{i_{1}\ldots i_{6}}K_{i_{1}}^{j_{1}}\d_{i_{2}}^{j_{2}}\Big(\frac34\car^{j_3j_4}{}_{i_3i_4}\car^{j_5j_6}{}_{i_5i_6}-\frac{3}{2}\car^{j_3j_4}{}_{i_3i_4}K^{j_5}_{i_5}K^{j_6}_{i_6}+\frac{1}{2\ell^2}\car^{j_3j_4}{}_{i_3i_4}\d^{j_5}_{i_5}\d^{j_6}_{i_6}\NO\\
&\;+K^{j_3}_{i_3}K^{j_4}_{i_4}K^{j_5}_{i_5}K^{j_6}_{i_6}-\frac{2}{3\ell^2}K^{j_3}_{i_3}K^{j_4}_{i_4}\d^{j_5}_{i_5}\d^{j_6}_{i_6}+\frac{1}{5\ell^4}\d^{j_3}_{i_3}\d^{j_4}_{i_4}\d^{j_5}_{i_5}\d^{j_6}_{i_6}\Big).
\eal 

These can be written in more explicit form by carrying out the contractions of the generalized Kronecker delta with all tensor structures. This leads to the Kounterterm densities  
\bal\label{Kounterterms-expaned-low-d}
\cl\sbtx{K}^{d=2}=&\;-\frac{1}{2\k^2}\sqrt{-h}\,K,\NO\\
\cl\sbtx{K}^{d=3}=&\;\frac{\ell^2}{2\k^2}\sqrt{-h}\,\Big(K\car-2K^i_j\car^j_i-\frac13K^3+KK^i_jK^j_i-\frac23K^i_jK^j_kK^k_i\Big)-\frac{1}{\k^2}\sqrt{-h}\,K,\NO\\
\cl\sbtx{K}^{d=4}=&\;-\frac{3\ell^2}{16\k^2}\sqrt{-h}\,\Big(\frac{2}{3}K\car-\frac{4}{3}K^i_j\car^j_i-\frac{1}{3}K^3+KK^i_jK^j_i-\frac{2}{3}K^i_jK^j_kK^k_i\Big)-\frac{9}{8\k^2}\sqrt{-h}\,K,\NO\\
\cl\sbtx{K}^{d=5}=&\;\frac{\ell^4}{6\k^2}\sqrt{-h}\,\Big(\frac{1}{4}K\car^{ij}{}_{kl}\car^{kl}{}_{ij}-KK^k_iK^l_j\car^{ij}{}_{kl}-K_{ij}\car^{ikpq}\car^{j}{}_{kpq}+2K^i_kK^{kj}K^{pq}\car_{ipjq}\NO\\
&\;+2K^{ij}\car^{kl}\car_{ikjl}+\frac{1}{4}K\car^2-K^{ij}\car_{ij}\car-\frac{1}{6}K^3\car+\frac{1}{2}KK^i_jK^j_i\car-\frac{1}{3}K^i_jK^j_kK^k_i\car\NO\\
&\;-K\car^i_j\car^j_i+K^2K^i_j\car^j_i-K^i_jK^j_iK^k_l\car^l_k+2K^i_j\car^k_i\car_k^j-2KK^i_jK^j_k\car^k_i+2K^i_jK^j_kK^k_l\car^l_i\NO\\
&\;+\frac{1}{20}K^5-\frac{1}{2}K^3K^i_jK^j_i+\frac{3}{4}K(K^i_jK^j_i)^2+K^2K^i_jK^j_kK^k_i-K^i_jK^j_iK^l_pK^p_qK^q_l\NO\\
&\;-\frac{3}{2}KK^i_jK^j_kK^k_lK^l_i+\frac{6}{5}K^i_jK^j_kK^k_lK^l_pK^p_i\Big)-\frac{1}{\k^2}\sqrt{-h}\,K,\NO\\
\cl\sbtx{K}^{d=6}=&\;\frac{5\ell^4}{12\times8\k^2}\sqrt{-h}\,\Big(\frac{3}{20}K\car^{ij}{}_{kl}\car^{kl}{}_{ij}-\frac{9}{10}KK^k_iK^l_j\car^{ij}{}_{kl}-\frac{3}{5}K_{ij}\car^{ikpq}\car^{j}{}_{kpq}+\frac{9}{5}K^i_kK^{kj}K^{pq}\car_{ipjq}\NO\\
&\;+\frac{6}{5}K^{ij}\car^{kl}\car_{ikjl}+\frac{3}{20}K\car^2-\frac{3}{5}K^{ij}\car_{ij}\car-\frac{3}{20}K^3\car+\frac{9}{20}KK^i_jK^j_i\car-\frac{3}{10}K^i_jK^j_kK^k_i\car\NO\\
&\;-\frac{3}{5}K\car^i_j\car^j_i+\frac{9}{10}K^2K^i_j\car^j_i-\frac{9}{10}K^i_jK^j_iK^k_l\car^l_k+\frac{6}{5}K^i_j\car^k_i\car_k^j-\frac{9}{5}KK^i_jK^j_k\car^k_i+\frac{9}{5}K^i_jK^j_kK^k_l\car^l_i\NO\\
&\;+\frac{1}{20}K^5-\frac{1}{2}K^3K^i_jK^j_i+\frac{3}{4}K(K^i_jK^j_i)^2+K^2K^i_jK^j_kK^k_i-K^i_jK^j_iK^l_pK^p_qK^q_l-\frac{3}{2}KK^i_jK^j_kK^k_lK^l_i\NO\\
&\;+\frac{6}{5}K^i_jK^j_kK^k_lK^l_pK^p_i-\frac{2}{5\ell^2}K^i_jK^j_kK^k_i+\frac{3}{5\ell^2}KK^i_jK^j_i-\frac{1}{5\ell^2}K^3-\frac{3}{5\ell^2}K^i_j\car^j_i+\frac{3}{10\ell^2}K\car\Big)\NO\\
&\;-\frac{15}{16\k^2}\sqrt{-h}\,K.
\eal 
 
The last and most tedious step is to insert the covariant expansion of the extrinsic curvature \eqref{K-exp} in \eqref{Kounterterms-expaned-low-d} and keep terms of dilatation weight up to (and including) zero, i.e. up to asymptotically finite terms. Up to $d=6$ the result is
\bal\label{Kounterterms-R-low-d}
\cl\sbtx{K}^{d=2}=&\;-\frac{1}{\k^2\ell}\sqrt{-h}\Big(1+\frac{\ell^2}{4}\car+\cdots\Big),\NO\\
\cl\sbtx{K}^{d=3}=&\;-\frac{1}{\k^2\ell}\sqrt{-h}\Big(2+\frac{\ell^2}{2}\car+\cdots\Big),\NO\\
\cl\sbtx{K}^{d=4}=&\;-\frac{1}{\k^2\ell}\sqrt{-h}\Big(3+\frac{\ell^2}{4}\car-\frac{\ell^4}{8}\big(\cp^{ij}\cp_{ij}-\cp^2\big)+\cdots\Big),\NO\\
\cl\sbtx{K}^{d=5}=&\;-\frac{1}{\k^2\ell}\sqrt{-h}\Big(4+\frac{\ell^2}{6}\car+\frac{\ell^4}{2}\Big(\cp^{ij}\cp_{ij}-\cp^2-\frac{1}{12}\cw_{ijkl}\cw^{ijkl}\Big)+\cdots\Big),\NO\\
\cl\sbtx{K}^{d=6}=&\;-\frac{1}{\k^2\ell}\sqrt{-h}\Big(5+\frac{\ell^2}{8}\car+\frac{\ell^4}{4}\Big(\cp^{ij}\cp_{ij}-\cp^2-\frac{1}{16}\cw^{ijkl}\cw_{ijkl}\Big)\\
&\;-\frac{\ell^6}{32}\Big(\frac{5}{3}\big(2\cp^i_j\cp^j_k\cp^k_i-3\cp\cp^i_j\cp^j_i+\cp^3\big)+\cp^{ij}\cp^{kl}\cw_{ikjl}-\Big(\cp^{ij}-\frac14\cp h^{ij}\Big)\cw_{ikpq}\cw_j{}^{kpq}\Big)+\cdots\Big),\NO
\eal 
where the ellipses stand for covariant terms of negative dilatation weight, i.e. terms that asymptotically vanish. As advertised, the expressions \eqref{Kounterterms-R-low-d} for the Kounterterms involve only the intrinsic curvature of the induced metric and can therefore be compared directly with the boundary counterterms \eqref{counterterms}, whose explicit form up to $d=6$ is 
\bal\label{counterterms-low-d}
\cl\sbtx{ct}^{d=2}=&\;-\frac{1}{\k^2\ell}\sqrt{-h}\Big(1-\frac{\ell^2}{4}\log(e^{-2r_c/\ell})\car+s_0\car\Big),\NO\\
\cl\sbtx{ct}^{d=3}=&\;-\frac{1}{\k^2\ell}\sqrt{-h}\Big(2+\frac{\ell^2}{2}\car\Big),\NO\\
\cl\sbtx{ct}^{d=4}=&\;-\frac{1}{\k^2\ell}\sqrt{-h}\Big(3+\frac{\ell^2}{4}\car-\frac{\ell^4}{4}\log(e^{-2r_c/\ell})\big(\cp^{ij}\cp_{ij}-\cp^2\big)\Big)+s_0\ce_4+s_1 \cw_{ijkl}\cw^{ijkl},\NO\\
\cl\sbtx{ct}^{d=5}=&\;-\frac{1}{\k^2\ell}\sqrt{-h}\Big(4+\frac{\ell^2}{6}\car+\frac{\ell^4}{2}\big(\cp^{ij}\cp_{ij}-\cp^2\big)\Big),\NO\\
\cl\sbtx{ct}^{d=6}=&\;-\frac{1}{\k^2\ell}\sqrt{-h}\Big(5+\frac{\ell^2}{8}\car+\frac{\ell^4}{4}\big(\cp^{ij}\cp_{ij}-\cp^2\big)\\
&\;-\frac{\ell^6}{32}\log(e^{-2r_c/\ell})\big(\cp_{ij}\cb^{ij}+2\cp^i_j\cp^j_k\cp^k_i-3\cp\cp^i_j\cp^j_i+\cp^3\big)\Big)+s_0\ce_6+s_1\ci_1+s_2\ci_2+s_3\ci_3.\NO
\eal
As discussed in section \ref{scheme}, the arbitrary constants $s_0,\,s_1,\,\ldots$ parameterize the general form of the finite local counterterms that correspond to the renormalization scheme dependence of the dual field theory in even dimensions. 

Comparing the expressions \eqref{Kounterterms-R-low-d} and \eqref{counterterms-low-d} immediately leads to a few general conclusions. Firstly, it is clear that the only dimension for which the Kounterterms agree fully with the boundary counterterms, and hence regularize the variational problem for general AlAdS manifolds, is $d=3$ (i.e. AdS$_4$). For no other dimension do the Kounterterms provide the required boundary term for general AlAdS manifolds. A universal divergence that is not canceled by the Kounterterms is the logarithmic divergence in even dimensions $d$ (odd bulk). Moreover, for both even and odd $d\geq 5$, power law divergences in the Kounterterms also differ from those in the counterterms by terms involving the Weyl tensor of the induced metric. Finally, the Kounterterms give rise to specific local and covariant finite terms for even $d$, corresponding to a specific choice of renormalization scheme. However, starting with $d=6$, these finite terms are in general not a sum of global and local conformal invariants, as is the case for the boundary counterterms.  
   
Although the Kounterterms provide the required boundary term for general AlAdS manifolds only in bulk dimension four, it is possible that in other dimensions they agree with the boundary counterterms on a restricted class of AlAdS manifolds. From the above comparison follows that a necessary condition for such an agreement is that the Weyl tensor of the boundary metric vanishes. In the case of odd dimensions, an additional requirement is that the $Q$-curvature, i.e. the conformal anomaly, is also zero. Since for conformally flat manifolds all local Weyl invariants are zero, the additional condition for odd dimensional AlAdS manifolds is equivalent to the vanishing of the Euler-Poincar\'e density. In the next section we will show that these conditions are also sufficient. 

It must be stressed, however, that agreement between the Kounterterms and counterterms on a restricted class of AlAdS backgrounds does not automatically ensure that quantities such as conserved charges, or higher-point holographic correlation functions, are renormalized by the Kounterterms. This is because successive derivatives of the Kounterterms with respect to the induced metric need not agree with the corresponding quantity obtained from the boundary counterterms. In the next section we will demonstrate that for AlAdS manifolds with a conformally flat boundary, the agreement persists at least for the canonical momenta, i.e. for holographic one-point functions. Agreement for higher-point functions is guaranteed only when the Kounterterms coincide with the counterterms for arbitrary AlAdS manifolds, i.e. only in four dimensions.

\section{AlAdS manifolds with conformally flat boundary}
\label{sec:conflat}

In the previous section we saw that, except in four dimensions ($d=3$), the Kounterterms regularize the variational problem of AdS gravity only within a subclass of AlAdS manifolds that have a vanishing boundary Weyl tensor and (in the case of odd bulk dimension) all logarithmic divergences are numerically zero. For $d>3$, an AlAdS$_{d+1}$ manifold with a vanishing boundary Weyl tensor is necessarily asymptotically conformally flat, i.e. the bulk Weyl tensor is zero, up to possible contributions from the normalizable mode only. This can be shown using the leading asymptotic form of the components of the bulk Weyl tensor in \eqref{Weyl-asymptotics} as follows. 

Firstly, it is manifest from the relations \eqref{Weyl-asymptotics} that a vanishing bulk Weyl tensor implies that the boundary Weyl, Cotton and Bach tensors vanish, and so the boundary is conformally flat. The converse is not necessarily true \cite{Skenderis:1999nb,deHaro:2000xn}, but it does hold, up to contributions due to the normalizable mode. In particular, suppose that the boundary Weyl tensor vanishes. For $d>3$, this implies that the Cotton and Bach tensors of the boundary metric also vanish, and hence the leading asymptotic form of the bulk Weyl tensor is zero due to the relations \eqref{Weyl-asymptotics}. However, two AlAdS manifolds with the same boundary metric can only differ in the normalizable mode of the bulk metric and, therefore, the bulk Weyl tensor must vanish, up to possible normalizable contributions.

The Kounterterms, therefore, may potentially regularize the variational problem for AdS$_{d+1}$ gravity when $d>3$ only within the subclass of asymptotically conformally flat AlAdS manifolds, i.e. those with vanishing bulk Weyl tensor, up to possible normalizable contributions. However, generic odd dimensional asymptotically conformally flat AlAdS manifolds still have a logarithmic divergence and so additional conditions on the boundary metric must be imposed in that case. The relevant condition is that Branson's $Q$-curvature is also zero, up to a trivial total derivative. From the decomposition \eqref{Q-curvature} of the $Q$-curvature, it follows that, for conformally flat manifolds, the $Q$-curvature coincides (up to a globally defined total divergence) with the Euler-Poincar\'e density since, all local conformal invariants vanish. The additional condition for odd dimensional AlAdS manifolds, therefore, amounts to demanding that the Euler characteristic of the boundary is zero.

In this section, we show that these conditions, summarized in table~\ref{conditions} in the introduction, are not only necessary for the Kounterterms to regularize the AdS variational problem, but also sufficient. To this end, we first determine the form of the boundary counterterms for asymptotically conformally flat AlAdS manifolds of arbitrary dimension. We then compare these with the Kounterterms for AlAdS manifolds subject to the conditions given in table~\ref{conditions}.

\subsection{Counterterms for asymptotically conformally flat AlAdS manifolds} 

We refer to AlAdS manifolds with a conformally flat boundary as {\em asymptotically conformally flat}. 
For AlAdS$_{d+1}$ manifolds with boundary dimension $d>2$, asymptotic conformal flatness is equivalent to the vanishing of the boundary Weyl, Cotton and Bach tensors.\footnote{For $d=3$, the vanishing of the boundary Weyl tensor holds for any metric, but the vanishing of the Cotton and Bach tensors is still non trivial.} This in turn implies that the bulk Weyl tensor is asymptotically zero, except for possible contributions from the normalizable mode of the bulk metric. Notice that any AlAdS$_3$ ($d=2$) manifold is asymptotically conformally flat since any two dimensional boundary is conformally flat. This is reflected in the fact that the bulk Weyl tensor vanishes identically, while the Cotton tensor is zero due to the Einstein condition. 

Since the normalizable mode of the bulk metric does not contribute to the long distance divergences of the on-shell action, the boundary counterterms for asymptotically conformally flat AlAdS manifolds are identical to those for conformally flat ones, for which the bulk Weyl tensor is identically zero. For the purpose of determining the boundary counterterms for asymptotically conformally flat AlAdS manifolds therefore, it suffices to consider strictly conformally flat ones. 

Setting the bulk Weyl tensor to zero and using Einstein's equations leads to the three conditions (see eq.~\eqref{Weyl-on-shell} in appendix \ref{RadialFoliation})  
\bal\label{ConFlatEqs}
&\dot{K}^i_j+K^i_kK^k_j-\frac{1}{\ell^2}\d^i_j=0,\NO\\
&D_kK^i_j-D_jK^i_{k}=0,\NO\\
&\car^{ik}{}_{jl}-K^i_jK^k_l+K^i_lK^k_j-\frac{1}{\ell^2}\d^i_l \d^k_j+\frac{1}{\ell^2}\d^i_j \d^k_l=0.
\eal
An immediate geometric implication of these equations is that the Cotton and Weyl tensors of the induced metric $h_{ij}$ vanish, namely
\bal\label{ConFlatGeom}
\cc_{ijk}=&\;D_k\cp_{ij}-D_j\cp_{ik}=0,\NO\\
\cw_{ikjl}=&\;\car_{ikjl}+h_{il}\cp_{kj}+h_{kj}\cp_{il}-h_{ij}\cp_{kl}-h_{kl}\cp_{ij}=0.
\eal
These equations correspond to the leading order terms of respectively the second and third equations in \eqref{ConFlatEqs}, when expanded covariantly in eigenfunctions of the dilatation operator (see \eqref{Weyl-asymptotics}). 

Combining the third equation in \eqref{ConFlatEqs} and second one in \eqref{ConFlatGeom} results in yet another identity relating the extrinsic curvature and the Schouten tensor algebraically, namely
\be\label{K-Schouten}
K^i_jK^k_l-K^i_lK^k_j=\d^i_{j}\cp^k_{l}+\d^k_{l}\cp^i_{j}-\d^i_{l}\cp^k_{j}-\d^k_{j}\cp^i_{l}+\frac{1}{\ell^2}\d^i_j \d^k_l-\frac{1}{\ell^2}\d^i_l \d^k_j.
\ee   
Using its traces, this equation can be recast as an algebraic condition on $K^i_j$ only, 
\bal\label{K-traces}
K^i_jK^k_l-K^i_lK^k_j=&\;\frac{1}{d-2}\Big[\d^i_j\big(K^k_lK-K^k_pK^p_l\big)+\d^k_l\big(K^i_jK-K^i_pK^p_j\big)-\d^i_l\big(K^k_jK-K^k_pK^p_j\big)\NO\\
&\;\hskip2.0cm-\d^k_j\big(K^i_lK-K^i_pK^p_l\big)-\frac{1}{d-1}(K^2-K^p_qK^q_p)\big(\d^i_j\d^k_l-\d^i_l\d^k_j\big)\Big].
\eal
A more useful form of this equation is  
\be\label{SymId}
Y^{i_1i_2|i_3i_4}_{\,j_1j_2|j_3j_4}K^{j_3}_{i_3}K^{j_4}_{i_4}=0,
\ee
where 
\be\label{projector}
Y^{i_1i_2|i_3i_4}_{\,j_1j_2|j_3j_4}\equiv \frac{1}{4}\Big(\d^{i_1i_2i_3i_4}_{j_1j_2j_3j_4}-\frac{d-3}{d-2}\big(\d^{i_1}_{j_1}\d^{i_2i_3i_4}_{j_2j_3j_4}+\d^{i_2}_{j_2}\d^{i_1i_3i_4}_{j_1j_3j_4}-\d^{i_1}_{j_2}\d^{i_2i_3i_4}_{j_1j_3j_4}-\d^{i_2}_{j_1}\d^{i_1i_3i_4}_{j_2j_3j_4}\big)+\frac{d-3}{d-1}\d^{i_1i_2}_{j_1j_2}\d^{i_3i_4}_{j_3j_4}\Big),
\ee
is a projection operator that projects onto the traceless part of rank 4 tensors with the symmetries of the Riemann tensor. Namely, it annihilates any tensor of the form $\d^{[i}_{j}\cm^{k]}_{l}+\cm^{[i}_{j}\d^{k]}_{l}$. For example, it projects the Riemann tensor to its Weyl part: 
\be
Y_{\,j_{1}j_{2}|j_{3}j_{4}}^{i_{1}i_{2}|i_{3}i_{4}}\car^{j_3j_4}{}_{i_3i_4}=\cw^{i_1i_2}{}_{j_1j_2}\,.
\ee

Inserting the expansion of the extrinsic curvature in eigenfunctions of the dilation operator in \eqref{SymId} results in an identical equation for the Schouten tensor of the induced metric, i.e.  
\be\label{SymIdSchouten}
Y^{i_1i_2|i_3i_4}_{\,j_1j_2|j_3j_4}\cp^{j_3}_{i_3}\cp^{j_4}_{i_4}=0. 
\ee
Hence, both the extrinsic curvature and the Schouten tensor obey the same algebraic constraints, which play an important role in the subsequent analysis.

\subsubsection*{Fefferman-Graham expansion} 

The first equation in \eqref{ConFlatEqs} can be integrated to obtain the exact form of the bulk metric. Writing the induced metric and extrinsic curvature in matrix notation as $(\bb h)^i_j=h_{ij}$, $(\bb K)^i_j=K^i_j$ and inserting the defining relation 
\be
\bb K=\frac{1}{2}\pa_r \log \bb h,
\ee
in \eqref{ConFlatEqs}, one finds that the Fefferman-Graham expansion for conformally flat AlAdS manifolds terminates. The exact form of the induced metric is \cite{Skenderis:1999nb,deHaro:2000xn}
\be
h_{ij}=e^{2r/\ell}\big(g_{(0)ij}(x)+e^{-2r/\ell}g_{(2)ij}(x)+e^{-4r/\ell}g_{(4)ij}(x)\big),
\ee 
where $g\sub{0}_{ij}$ is a conformally flat boundary metric, $g_{(2)ij}=-\ell^2\cp_{ij}[g_{(0)}]$ for $d>2$, and $g_{(4)ij}=(g_{(2)}g^{-1}_{(0)}g_{(2)})_{ij}/4$. In the case of two dimensional boundary, $g_{(2)ij}$ determines the boundary stress tensor and is arbitrary, except for a divergence and a trace constraint.

\subsubsection*{On-shell action in terms of the extrinsic curvature} 

A remarkable consequence of equations \eqref{ConFlatEqs} for conformally flat AlAdS manifolds is that they allow us to obtain an exact expression for the on-shell action in the case of odd $d$ in terms of the extrinsic curvature. Evaluating the bulk radial Lagrangian \eqref{rLagrangian} on-shell gives
\be\label{HJdensity-eq}
\mathscr{L}=\dot\cl=\frac{1}{\k^2}\sqrt{-h}\big(K^2-K^i_jK^j_i\big),
\ee
where recall that $\cl$ is the Hamilton-Jacobi density defined in \eqref{HJdensity}. This identity holds for any solution of the bulk field equations, but we will now show that, for conformally flat manifolds, the first equation in \eqref{ConFlatEqs} allows us to integrate \eqref{HJdensity-eq} and determine $\cl$ exactly.

To this end, it is necessary to introduce the symmetric polynomials of the matrix $(\bb K)^i_j=K^i_j$
\be\label{sigma-K}
\s_k(\bb K)\equiv\frac{1}{(d-k)!k!}\d^{j_1j_2\cdots j_d}_{i_1i_2\cdots i_d}K^{i_1}_{j_1}K^{i_2}_{j_2}\cdots K^{i_k}_{j_k}\d^{i_{k+1}}_{j_{k+1}}\cdots \d^{i_{d}}_{j_{d}}=\frac{1}{k!}\d^{j_1j_2\cdots j_k}_{i_1i_2\cdots i_k}K^{i_1}_{j_1}K^{i_2}_{j_2}\cdots K^{i_k}_{j_k},
\ee
where the generalized Kronecker delta was defined in \eqref{antisym-deltas} and $\s_{k}(\bb P)=0$ for $k>d$. A brief review of symmetric polynomials in the context of conformal geometry can be found in \cite{10.1093/imrn/rnt095} and in appendix \ref{SymmTr} we summarize the properties most relevant to our analysis. Notice that equation \eqref{HJdensity-eq} for the Hamilton-Jacobi density can be expressed as 
\be\label{HJdensity-eq-sigma}
\dot\cl=\frac{2}{\k^2}\sqrt{-h}\,\s_2(\bb K).
\ee

Using the first equation in \eqref{ConFlatEqs}, one can show that the symmetric polynomials of the extrinsic curvature satisfy the recursion relation
\be\label{sigma-recursion}
\pa_r\big(\sqrt{-h}\,\s_k(\bb K)\big)=\sqrt{-h}\big((d-k+1)\ell^{-2}\s_{k-1}(\bb K)+(k+1)\s_{k+1}(\bb K)\big).
\ee
Notice that the r.h.s. of this relation involves symmetric polynomials of either even or odd order. Given the form \eqref{HJdensity-eq-sigma} of the equation for $\cl$, this motivates us to look for a solution of the form
\be\label{HJansatz}
\cl=\frac{\sqrt{-h}}{\k^2\ell}\sum_{k=0}^{\big[\frac{d-1}{2}\big]}\a_k\;\s_{2k+1}(\ell\bb K),
\ee
where $\a_k$ are coefficients to be determined. However, the relevant solution for $\cl$ must have the correct asymptotic behavior, which amounts to the condition (see \eqref{L0})
\be\label{HJbc}
\sum_{k=0}^{\big[\frac{d-1}{2}\big]}\a_k\;\s_{2k+1}(\bb 1)=\sum_{k=0}^{\big[\frac{d-1}{2}\big]}\left(\begin{matrix} d\\ 2k+1\end{matrix}\right)\a_k=d-1.
\ee

It is straightforward to check that a solution of the form \eqref{HJansatz} satisfying the condition \eqref{HJbc} exists only for odd $d$ (even bulk) and takes the form 
\be\label{HJ-sol-ConFlat-even}
\cl=-\frac{\sqrt{-h}}{\k^2\ell\G\big(\frac{d}{2}\big)}\sum_{k=1}^{\big[\frac{d-1}{2}\big]} (-1)^k \G(k+1)\G\Big(\frac{d}{2}-k\Big)\s_{2k+1}(\ell\bb K),\qquad d=2n-1,\quad n\geq 2.
\ee
This is an exact solution of the radial Hamilton-Jacobi equation for even dimensional conformally flat AlAdS manifolds. The corresponding solution for odd dimensional conformally flat AlAdS manifolds cannot be expressed as a polynomial in the extrinsic curvature. However, we will see that the asymptotic form of the on-shell action for both even and odd dimensions, up to the relevant order in the dilatation operator expansion, can be deduced directly from the exact solution \eqref{HJ-sol-ConFlat-even}, once the extrinsic curvature is expressed in terms of the Schouten tensor of the induced metric. 

From the exact solution \eqref{HJ-sol-ConFlat-even} of the Hamilton-Jacobi equation, we conclude that the boundary counterterms for even dimensional asymptotically conformally flat AlAdS manifolds take the form\footnote{Notice that this expression is not manifestly local in boundary derivatives -- as the counterterms must be -- since it depends on the extrinsic curvature. However, it does turn out to be local, once the explicit form of extrinsic curvature as a function of the induced metric is taken into account, as we will verify below.} 
\be\label{Counterterms-ConFlat-even}
\cl\sbtx{ct}=\frac{\sqrt{-h}}{\k^2\ell\G\big(\frac{d}{2}\big)}\sum_{k=1}^{\big[\frac{d-1}{2}\big]} (-1)^k \G(k+1)\G\Big(\frac{d}{2}-k\Big)\s_{2k+1}(\ell\bb K),\qquad d=2n-1.
\ee
In the next subsection, we will show that this expression coincides with the boundary Kounterterms for even dimensional asymptotically conformally flat AlAdS manifolds.

\subsubsection*{Extrinsic curvature in terms of the Schouten tensor}  

We have found an exact solution of the Hamilton-Jacobi equation for even dimensional conformally flat AlAdS manifolds in terms of the extrinsic curvature $K^i_j$. However, in order to determine the explicit form of the boundary counterterms (and verify that they are local) it is necessary to also evaluate the extrinsic curvature as a function of the induced metric. As we will show, when expressed in terms of the induced metric and generic boundary dimension $d$, the counterterms take identical form for even and odd asymptotically conformally flat AlAdS manifolds.

A significant simplification in the case of conformally flat AlAdS manifolds is that eq.~\eqref{K-Schouten} determines the extrinsic curvature {\em algebraically} in terms of the Schouten tensor, $\cp^i_j$, of the induced metric. However, the relation between the on-shell action (Hamilton-Jacobi functional) and the extrinsic curvature (equivalently the canonical momentum) is less clear once we restrict to conformally flat metrics, since generically we should expect that
\be\label{ConflatMomentum-def}
\left.\p^{ij}\right|\sbtx{Conf.\,Flat}=\left[\frac{\d}{\d h_{ij}}\int_{\pa\cm_{r_c}}\hskip-0.2cm\tx d^dx\,\cl\right]_{\text{Conf.\,Flat}}\neq\frac{\d}{\d h_{ij}}\int_{\pa\cm_{r_c}}\hskip-0.2cm\tx d^dx\left.\cl\right|\sbtx{Conf.\,Flat}\equiv \P^{ij}.
\ee
In particular, the recursive algorithm for determining $\p^{ij}$ and $\cl$ in tandem discussed in section \ref{sec:hr} does not necessarily apply once we restrict to conformally flat manifolds. This is not to say that the algorithm definitely does not apply, but merely that we should not assume that it does. It may or may not apply, and we need to address this question by evaluating both the on-shell action and the canonical momentum in an independent way. 

Conformal flatness implies that the Hamilton-Jacobi density can be parameterized as
\be\label{HJ-F}
\left.\cl\right|\sbtx{Conf.\,Flat}=\sqrt{-h}\,\cf(\bb P),
\ee
where $\cf(\bb P)$ is a yet unspecified scalar function that admits a Taylor expansion in the Schouten tensor of the induced metric. $\P^{ij}$ on the r.h.s. of \eqref{ConflatMomentum-def} can be evaluated in terms of the tensor 
\be
\ct^i_j\equiv\frac{\pa\cf}{\pa \cp^j_i}.
\ee
A small calculation shows that 
\bal\label{ConflatMomentum}
\P^{ij}=&\;\sqrt{-h}\Big\{\frac12h^{ij}\cf-\frac{1}{2(d-2)}\Big[(d-4)\ct^{(i}_k\cp^{kj)}+\cp\ct^{ij}+\ct^k_l\cp^l_k h^{ij}+\frac{1}{d-1}\ct(\cp^{ij}-\cp h^{ij})\NO\\
&\;-D^{(i}D^k\ct^{j)}_k+h^{ij}D^kD_l\ct^l_k+D_k\big(D^{[k}\ck^{i]j}+D^{[k}\ck^{j]i}\big)\Big]\Big\},
\eal
where
\be
\ck^{ij}\equiv\ct^{ij}-\frac{1}{d-1}\ct h^{ij}.
\ee
Diffeomorphism invariance along the radial slice implies the two conservation equations 
\be\label{ConfConservations}
D_i\P^i_j=0,\qquad D_k(\ct^k_l\cp^l_i-\cp^k_l\ct^l_i)+(D_i\cp^l_k)\ct^k_l-D_i\cf=0,
\ee
which hold independently of the specific form of $\cf$. 

From \eqref{ConflatMomentum}, it follows that a sufficient condition for $\P^{ij}$ to be algebraic in terms of the Schouten tensor (and hence potentially agree with $\p^{ij}$) is that $\ct^i_j$ satisfies 
\be\label{ConfConstraint}
D_{[i}\ct^k_{j]}-\frac{1}{d-1}D_{[i}\ct\d^k_{j]}=0 \Leftrightarrow D_{[i}\ck^k_{j]}=0,
\ee
which also implies that 
\be
D_i\ct^i_j=0.
\ee
This ensures that all derivative terms in \eqref{ConflatMomentum} vanish and $\P^{ij}$ is algebraic in terms of $\cp^i_j$. Moreover, since the expansion of $\cf(\bb P)$ in eigenfunctions of the dilatation operator involves a sum of homogeneous polynomials in $\cp^i_j$ that satisfy the identity
\be
\ct_{(2n)}{}^i_j\cp^j_i=n\cf_{(2n)},\qquad n\geq1,
\ee
the constraint \eqref{ConfConstraint} implies that the trace of \eqref{ConflatMomentum} takes the form
\be
\P_{(2n)}=\Big(\frac{d-2n}{2}\Big)\sqrt{-h}\,\cf_{(2n)},
\ee
which is the same as the identity \eqref{trace-relations} that $\p^{ij}$ satisfies for generic AlAdS manifolds. This provides further evidence that the constraint \eqref{ConfConstraint} is the key to answering the question whether $\P^{ij}$ agrees with $\p^{ij}$ in the case of conformally flat AlAdS manifolds.  

As we reviewed in section \ref{sec:hr}, the $Q$-curvature for odd dimensional AlAdS manifolds can be decomposed into a sum of the Euler-Poincar\'e density of the induced metric and a local conformal invariant, which vanishes for conformally flat manifolds. Moreover, it is straightforward to show that the (generalized -- i.e. $2n\leq d$) Euler-Poincar\'e density, $\ce_{2n}$, of a conformally flat metric reduces to a symmetric polynomial of its Schouten tensor, namely $\ce_{2n}\propto \s_n(\bb P)$ (see Proposition 2.2 in \cite{10.1093/imrn/rnt095}). Since the boundary counterterms can be thought of as the sum of $Q$-curvatures in all even boundary dimensions, one may expect that the function $\cf$ that parameterizes the Hamilton-Jacobi functional for conformally flat manifolds is a sum of symmetric polynomials of the Schouten tensor, i.e. $\cf_{(2n)}\propto \s_n(\bb P)$. We will now evaluate $\cf_{(2n)}$ for any $n$ and confirm that this is indeed the case. 

The key to determining the polynomials $\cf_{(2n)}$ is equation \eqref{SymIdSchouten}, which  implies a number of algebraic relations among symmetric polynomials of the Schouten tensor of conformally flat manifolds. Recall that the $k$-th symmetric polynomial of the Schouten tensor is given by (see appendix \ref{SymmTr})
\be\label{sigma-P}
\s_k(\bb P)=\frac{1}{k!}\d^{j_1j_2\cdots j_k}_{i_1i_2\cdots i_k}\cp^{i_1}_{j_1}\cp^{i_2}_{j_2}\cdots \cp^{i_k}_{j_k}.
\ee
In the case of the Schouten tensor, $\s_k(\bb P)$ is known as the $k$-th order Meissner-Olechowski density \cite{Meissner:2000dy}. The $k$-th Newton transformation of the Schouten tensor is defined as
\be\label{k-th-Newton-P}
(\bb T_k(\bb P))^i_j\equiv\frac{\pa}{\pa \cp^j_i}\s_{k+1}(\bb P)=\frac{1}{k!}\d^{ii_2\cdots i_{k+1}}_{jj_2\cdots j_{k+1}}\cp^{j_2}_{i_2}\cdots \cp^{j_{k+1}}_{i_{k+1}}.
\ee
The symmetric polynomials and the associated Newton transform can be defined for any $d\times d$ matrix, but for the Schouten tensor of conformally flat manifolds these objects have a much richer structure. For example, the vanishing of the Cotton tensor of the induced metric, $h_{ij}$, (see \eqref{ConFlatGeom}) implies that $\bb T_k(\bb P)$ are covariantly conserved for all $k$ (see e.g. Proposition 2.3 in \cite{10.1093/imrn/rnt095})
\be\label{CovDiv-T}
D_i(\bb T_k(\bb P))^i_j=\frac{1}{(k-1)!}\d^{ii_2\cdots i_{k+1}}_{jj_2\cdots j_{k+1}}(D_{[i}\cp^{j_2}_{i_2]})\cp^{j_3}_{i_3}\cdots \cp^{j_{k+1}}_{i_{k+1}}=0.
\ee

A number of less obvious properties of the symmetric polynomials and the associated Newton transform of the Schouten tensor of conformally flat manifolds follow from the algebraic constraint \eqref{SymIdSchouten}. As for the extrinsic curvature in \eqref{K-traces}, this constraint implies that the antisymmetrized tensor product of two Schouten tensors is determined by its traces, namely 
\be\label{P-traces}
\cp^i_{[j}\cp^k_{l]}=\frac{1}{d-2}\Big(\d^i_{[j}\big(\cp^k_{l]}\s_1-\cp^k_p\cp^p_{l]}\big)+\d^k_{[l}\big(\cp^i_{j]}\s_1-\cp^i_p\cp^p_{j]}\big)-\frac{2}{d-1}\s_2\d^i_{[j}\d^k_{l]}\Big).
\ee
Inserting this relation in the definition of the Newton transform of the Schouten tensor results in a number of algebraic identities, which we now derive.

Replacing a pair of Schouten tensors in \eqref{k-th-Newton-P} using \eqref{P-traces}, gives
\bal
(\bb T_k)^i_j=&\;\frac{1}{k!}\d^{ii_2\cdots i_{k+1}}_{jj_2\cdots j_{k+1}}\cp^{j_2}_{i_2}\cdots \cp^{j_{k+1}}_{i_{k+1}}\NO\\
=&\;\frac{2}{(d-2)}\frac{1}{k!}\d^{ii_2\cdots i_{k+1}}_{jj_2\cdots j_{k+1}}\cp^{j_2}_{i_2}\cdots \cp^{j_{k-1}}_{i_{k-1}}\Big(\d^{j_k}_{i_k}\big(\cp^{j_{k+1}}_{i_{k+1}}\s_1-\cp^{j_{k+1}}_p\cp^p_{i_{k+1}}\big)-\frac{1}{d-1}\s_2\d^{j_k}_{i_k}\d^{j_{k+1}}_{i_{k+1}}\Big)\\
=&\;\frac{2(d-k)}{k(d-2)}\Big(\s_1(\bb T_{k-1})^i_j-\frac{(d-k+1)}{(k-1)(d-1)}\s_2(\bb T_{k-2})^i_j-\frac{1}{(k-1)!}\d^{ii_2\cdots i_{k}}_{jj_2\cdots j_{k}}\cp^{j_2}_{i_2}\cdots \cp^{j_{k-1}}_{i_{k-1}}\cp^{j_k}_p\cp^{p}_{j_k}\Big),\NO
\eal
where we have used the identity
\be
\d^{i_1i_2\cdots i_{k}}_{j_1j_2\cdots j_{k}}\d^{j_k}_{i_k}=(d-k+1)\d^{i_1i_2\cdots i_{k-1}}_{j_1j_2\cdots j_{k-1}}.
\ee
Here and in the following we drop the arguments of $\bb T_k$ and $\s_k$ to simplify the notation, unless they are necessary for clarity. In order to evaluate the last term, we observe that
\bal
(k-1)\d^{ii_2\cdots i_{k}}_{jj_2\cdots j_{k}}\cp^{j_2}_{i_2}\cdots \cp^{j_{k-1}}_{i_{k-1}}\cp^{j_k}_p\cp^{p}_{j_k}=&\;(k-1)!\Big(\frac{\pa}{\pa\cp^j_i}\tr(\bb T_{k-1}\bb P^2)-(\bb T_{k-1}\bb P+\bb P\bb T_{k-1})^i_j\Big)\NO\\
=&\;(k-1)!\Big(\frac{\pa}{\pa\cp^j_i}\tr(\s_1\s_k-(k+1)\s_{k+1})+2(\bb T_{k}-\s_k\bb 1)^i_j\Big)\NO\\
=&\;(k-1)!\big(-\s_k\bb 1-(k-1)\bb T_k+\s_1\bb T_{k-1}\big)^i_j.
\eal
Hence, the symmetric polynomials of the Schouten tensor of a conformally flat manifold satisfy 
\be\label{id1}
\bb T_k=\frac{2(d-k)}{d(k-1)(k-2)}\Big((k-2)\s_1\bb T_{k-1}-\frac{d-k+1}{d-1}\s_2\bb T_{k-2}+\s_k\bb 1\Big),\qquad k>2.
\ee
The trace of this identity results in a recursion relation involving symmetric polynomials only
\be\label{id2}
\s_k=\frac{2(d-k+1)}{dk(k-3)}\Big((k-2)\s_1\s_{k-1}-\frac{d-k+2}{d-1}\s_2\s_{k-2}\Big),\qquad k>3.
\ee
These relations determine all $\bb T_k$ with $k>2$ in terms of $\bb T_2$, $\bb T_1$ and $\bb T_0=\bb 1$, as well as all $\s_k$ with $k>3$ in terms of $\s_3$, $\s_2$ and $\s_1$. 

However, $\bb T_2$ and $\s_3$ are also not independent. Combining \eqref{id1} with the general identity \eqref{T-id} leads to the two additional conditions
\be\label{id3}
\Big(\frac{d}{d-2}\t_3-\s_1\t_2\Big)^2+\frac{1}{d-1}\t_2^3=0,
\ee
and 
\be\label{id4}
\t_2\Big(\bb T_2-\frac{d-2}{d}\s_2\bb 1\Big)-\t_3\Big(\bb T_1-\frac{d-1}{d}\s_1\bb 1\Big)=0,
\ee
where
\be
\t_k\equiv dk\s_k-(d-k+1)\s_1\s_{k-1}.
\ee
It follows that $\bb T_k$ can be expressed as a linear combination of $\bb T_1$ and $\bb T_0=\bb 1$ for any $k\geq 2$. From  the recursion relations \eqref{id1} and \eqref{id2}, we determine that 
\be\label{TSol}
\t_2\bb T_k=\t_{k+1}\bb T_1-\frac{(k-1)(d-1)}{d-k-1}\t_{k+2}\bb 1,\qquad k\geq 2,
\ee
while $\t_k$ satisfy 
\be\label{tau-recursion}
\t_k=\frac{d-k+1}{d(d-1)(k-3)}\big((d-1)\s_1\t_{k-1}-(d-k+2)\s_{k-2}\t_2\big).
\ee

The identities derived above allow us to obtain a number of further results required for solving the conformal flatness equations \eqref{ConFlatEqs} and determining $K^i_j[h]$. The first is the recursion relation
\bal\label{id5}
&\s_l\bb T_{k-l}-\frac{d-k+l}{d-1}\s_{k-l}\bb T_l+\sum_{m=0}^{l-1}(\s_m\bb T_{k-m}-\s_{k-m}\bb T_{m})=\NO\\
&\frac{(k-1)!(d-k-1)!(d-2)!}{l!(k-l)!(d-k+l-1)!(d-l-1)!}\big((d(k-l)-k)\bb T_k-(d-k)(k-l)\s_k\bb 1\big),\quad 0\leq l\leq k,
\eal
which can be proved using \eqref{TSol}. The second result is a generalization of 
\eqref{P-traces} and follows directly from \eqref{P-traces} and the general form of $\bb T_k$ in \eqref{TSol}: 
\bal\label{P-traces-n}
&\Big((\bb T_{n})^i_{[j}-\frac{d-n}{d-1}\s_{n}\d^i_{[j}\Big)\cp^k_{l]}+\Big((\bb T_{n})^k_{[l}-\frac{d-n}{d-1}\s_{n}\d^k_{[l}\Big)\cp^i_{j]}=\NO\\
&\;\frac{n+1}{d-n-1}\big(\d^i_{[j}(\bb T_{n+1})^k_{l]}+\d^k_{[l}(\bb T_{n+1})^i_{j]}\big)-\frac{2(n+1)}{(d-1)}\s_{n+1}\d^i_{[j}\d^k_{l]}.
\eal

Finally, we can obtain a stronger version of the conservation equation \eqref{CovDiv-T}, namely
\be\label{T-Bianchi}
D_{[i}(\bb T_n)^k_{j]}-\frac{d-n}{d-1}D_{[i}\s_n \d^k_{j]}=0,\qquad n<d,\quad d>3,
\ee
which can be proved by induction as follows. Firstly, \eqref{T-Bianchi} holds for $n=1$ by virtue of the first equation in \eqref{ConFlatGeom}. We now show that if it holds for $n-1$, then it also holds for $n$. Adding zero to the identity \eqref{T-id} in the form $\b[\bb P,\bb T_{n-1}]$ for some arbitrary constant $\b$, we have 
\bal
2D_{[i}(\bb T_n)^k_{j]}=&\;2D_{[i}\s_n\d^k_{j]}-2D_{[i}(\b\bb P\bb T_{n-1}+(1-\b)\bb T_{n-1}\bb P)^k_{j]}\NO\\
=&\;2D_{[i}\s_n\d^k_{j]}-2\b(D_{[i}\cp^k_l)(\bb T_{n-1})^l_{j]}-2\b\cp^k_lD_{[i}(\bb T_{n-1})^l_{j]}-2(1-\b)\cp^l_{[j}D_{i]}(\bb T_{n-1})^k_{l}\NO\\
=&\;2D_{[i}\s_n\d^k_{j]}-\frac{2(d-n+1)}{d-1}D_{[i}\s_{n-1}\cp^k_{j]}+\frac{2(1-\b)(d-n+1)}{d-1}D_{l}\s_{n-1}\d^k_{[i}\cp^l_{j]}\NO\\
&-2\b D_l\big(\cp^k_{[i}(\bb T_{n-1})^l_{j]}\big)-2(1-\b)\cp^l_{[j}D_{l}(\bb T_{n-1})^k_{i]}\NO\\
=&\;2D_{[i}\s_n\d^k_{j]}-\frac{2(d-n+1)}{d-1}\big(D_{[i}\s_{n-1}\cp^k_{j]}+(1-\b)\s_{n-1}\d^k_{[i}D_{j]}\s_1\big)-2(1-\b)(D_{[i}\s_1)(\bb T_{n-1})^k_{j]}\NO\\
&-2 D_l\Big(\b\cp^k_{[i}(\bb T_{n-1})^l_{j]}+(1-\b)\cp^l_{[j}(\bb T_{n-1})^k_{i]}-\frac{(1-\b)(d-n+1)}{d-1}\s_{n-1}\d^k_{[i}\cp^l_{j]}\Big)\NO\\
=&\;2D_{[i}\Big(\s_n\d^k_{j]}+\frac{(d-n+1)}{d-1}\s_{n-1} ((\bb T_1)^k_{j]}-\b\s_1\d^k_{j]})-(1-\b)\s_1(\bb T_{n-1})^k_{j]}\Big)\NO\\
&-2 D_l\Big(\b\cp^k_{[i}(\bb T_{n-1})^l_{j]}+(1-\b)\cp^l_{[j}(\bb T_{n-1})^k_{i]}-\frac{(1-\b)(d-n+1)}{d-1}\s_{n-1}\d^k_{[i}\cp^l_{j]}\Big).
\eal
Setting $\b=1/2$, we can evaluate the last line using 
\eqref{P-traces-n}:
\bal
D_{[i}(\bb T_n)^k_{j]}=&\;D_{[i}\Big(\s_n\d^k_{j]}+\frac{(d-n+1)}{d-1}\s_{n-1} ((\bb T_1)^k_{j]}-\frac12\s_1\d^k_{j]})-\frac12\s_1(\bb T_{n-1})^k_{j]}\Big)\NO\\
&+D_l\Big(\frac{n}{d-n}\d^l_{[i}(\bb T_{n})^k_{j]}-\frac{2n}{(d-1)}\s_{n}\d^k_{[j}\d^l_{i]}+\frac{(d-n+1)}{d-1}\s_{n-1}\d^l_{[i}\cp^k_{j]}\Big)\NO\\
\stackrel{n\neq d}{=}&\;D_{[i}\Big(\s_n\d^k_{j]}-\frac{d(n-2)}{2(d-n)}\Big((\bb T_n)^k_{j]}-\frac{d-n}{d}\s_k\d^k_{j]}\Big)+\frac{n}{2(d-n)}(\bb T_{n})^k_{j]}-\frac{n}{d-1}\s_{n}\d^k_{[j}\Big),
\eal
where in the second equality we have used \eqref{id5} with $l=1$. Finally, collecting terms, we obtain
\be
\frac{n(d-3)}{2(d-n)}\Big(D_{[i}(\bb T_n)^k_{j]}-\frac{d-n}{d-1}D_{[i}\s_n \d^k_{j]}\Big)=0,
\ee
which completes the proof. We remark that all identities for the symmetric polynomials and Newton transform of the Schouten tensor derived in this section hold also for the extrinsic curvature, since they are a direct consequence of \eqref{SymIdSchouten}, which applies to the extrinsic curvature as well.

We now have the necessary tools to evaluate the extrinsic curvature as a function of the Schouten tensor for conformally flat AlAdS manifolds. This is determined by the algebraic matrix equation 
\be
(\tr\bb K) \bb K-\bb K^2=(\tr\bb P)\bb 1+(d-2)\bb P+\frac{d-1}{\ell^2}\bb 1,
\ee
obtained from a single index contraction of \eqref{K-Schouten}. Inserting a formal expansion of the extrinsic curvature in eigenfunctions of the dilation operator in this equation leads to the recursion relations
\bal\label{KRecursion}
&\bb K\sub{0}=\ell^{-1}\bb 1,\qquad \bb K\sub{2}=\ell\bb P,\NO\\
&(d-2)\bb K\sub{2k}+\tr\bb K\sub{2k}\bb 1=\ell\sum_{l=1}^{k-1} \big(\bb K\sub{2l}-\tr\bb K\sub{2l} \bb 1\big)\bb K\sub{2k-2l},\quad k>1.
\eal

The unique solution of these recursion relations is 
\be\label{KSol}\boxed{\boxed{
\bb K\sub{2k}-\tr\bb K\sub{2k}\bb 1=a_k\bb T_k(\bb P),\qquad a_k=\frac{(-1)^k\ell^{2k-1}(2k-2)!(d-k-1)!}{2^{k-1}(k-1)!(d-2)!},\qquad k\geq 1,}}
\ee
as can be readily checked using \eqref{id5}. In particular, the r.h.s. of \eqref{KRecursion} can be evaluated as follows:
\bal
&\;\ell\sum_{l=1}^{k-1} \big(\bb K\sub{2l}-\tr\bb K\sub{2l} \bb 1\big)\bb K\sub{2k-2l}=\ell\sum_{l=1}^{k-1} a_la_{k-l}\bb T_l\Big(\bb T_{k-l}-\frac{d-k+l}{d-1}\s_{k-l}\bb 1\Big)\NO\\
&\;=\ell\sum_{l=1}^{k-1} a_la_{k-l}\Big(\sum_{m=0}^{l-1}(\s_m\bb T_{k-m}-\s_{k-m}\bb T_{m})+\s_l\bb T_{k-l}-\frac{d-k+l}{d-1}\s_{k-l}\bb T_l\Big)\NO\\
&\;\stackrel{\eqref{id5}, \eqref{KSol}}{=}\frac{(-1)^k\ell^{2k-1}(k-1)!(d-k-1)!}{2^{k-2}(d-2)!}\times\NO\\
&\hspace{3.cm}\sum_{l=1}^{k-1} \frac{(2l-2)!(2k-2l-2)!}{l!(l-1)!(k-l)!(k-l-1)!}\big((d(k-l)-k)\bb T_k-(d-k)(k-l)\s_k\bb 1\big)\NO\\
&\;=a_k\big((d-2)\bb T_k-(d-k)\s_k\bb 1\big)=(d-2)\bb K\sub{2k}+\tr\bb K\sub{2k}\bb 1,
\eal
as required. Moreover, the identity \eqref{T-Bianchi} for $\bb T_k$ implies that the solution \eqref{KSol} satisfies the second equation in \eqref{ConFlatEqs} as well, and hence it is an exact solution of all equations \eqref{ConFlatEqs}. 

\subsubsection*{On-shell action in terms of the Schouten tensor} 

The main result in this section so far is the general solution \eqref{KSol} of the conformal flatness equations \eqref{ConFlatEqs}. Through the definition of the canonical momentum \eqref{momentum}, this determines
\be\label{momentum-sol}
\p\sub{0}^{ij}=\frac{(d-1)}{2\k^2\ell}\sqrt{-h}\;h^{ij},\qquad \p^{ij}_{(2k)}=-\frac{1}{2\k^2}a_k\sqrt{-h}\;(\bb T_k)^{ij},\quad k\geq 1.
\ee
Our next goal is to integrate these expressions to obtain the corresponding Hamilton-Jacobi functional in terms of the induced metric and its curvatures.

We saw earlier in this section that the Hamilton-Jacobi functional for conformally flat AlAdS manifolds can be parameterized in terms of a scalar function $\cf(\bb P)$ of the Schouten tensor as in \eqref{HJ-F}. Hence, we may ask what function $\cf(\bb P)$  the canonical momentum \eqref{momentum-sol} corresponds to. However, once we restrict to conformally flat metrics and parameterize the Hamilton-Jacobi functional in terms of the function $\cf(\bb P)$, its derivative with respect to the induced metric need not coincide with the canonical momentum $\p^{ij}$ in general, and so this question is a priori ill defined. 

A necessary condition for the derivative $\P^{ij}$ given in \eqref{ConflatMomentum} to agree with the canonical momentum is that the tensor $\ct^i_j$, obtained by differentiating $\cf(\bb P)$ with respect to $\cp^i_j$, satisfies the constraint \eqref{ConfConstraint}. Remarkably, identity \eqref{T-Bianchi} implies that the Newton transform $\bb T_k$ satisfies this constraint. This suggests that a suitable function $\cf(\bb P)$ corresponding to the canonical momentum \eqref{momentum-sol} does exist and can be expanded in eigenfunctions of the dilatation operator of the form
\be
\cf_{(2k)}(\bb P)=b_k\s_k(\bb P),
\ee
where $b_k$ are constants. It follows that, for $k\geq 1$,
\be
\ct_{(2k)j}^i=b_k(\bb T_{k-1})^i_j,
\ee
and so the derivative \eqref{ConflatMomentum} becomes
\be
\P^{ij}_{(2k)}=-\frac{b_k\sqrt{-h}}{2(d-2)}\Big(-(d-4)(\bb T_k)^{ij}+\s_1(\bb T_{k-1})^{ij}+(k-2)\s_k h^{ij}-\frac{d-k+1}{d-1}\s_{k-1}(\bb T_1)^{ij}\Big).
\ee
Using the identity \eqref{id5} for the case $l=1$, this reduces to 
\be
\P^{ij}_{(2k)}=b_k\sqrt{-h}\frac{(d-2k)}{2(d-k)}(\bb T_k)^{ij},
\ee
which coincides with the canonical momentum \eqref{momentum-sol}, provided 
\be
b_k=-\frac{(d-k)}{(d-2k)\k^2}a_k.
\ee
This shows that the Hamilton-Jacobi density corresponding to the canonical momentum \eqref{momentum-sol} is
\be\label{HJSol}\boxed{\boxed{
\cl_{(0)}=\frac{\sqrt{-h}}{\k^2}\frac{(d-1)}{\ell},\qquad \cl_{(2k)}=\frac{\sqrt{-h}}{\k^2}\frac{(-1)^{k+1}\ell^{2k-1}(2k-2)!(d-k)!}{2^{k-1}(k-1)!(d-2)!(d-2k)}\s_k(\bb P),\quad k\geq 1.}}
\ee

For odd $d$ (even bulk), \eqref{HJSol} is an exact solution of the Hamilton-Jacobi equation for conformally flat manifolds and is equivalent to the exact expression given in \eqref{HJ-sol-ConFlat-even} in terms of the extrinsic curvature. Nevertheless, the form \eqref{HJSol}  has two important advantages. Firstly, it confirms that for odd $d$ the exact solution of the Hamilton-Jacobi equation is a polynomial in the Schouten tensor, and hence a local functional of the induced metric. Secondly, it provides an {\em asymptotic} solution of the Hamilton-Jacobi equation for even $d$ as well. In particular, when $d$ is even, \eqref{HJSol} solves the recursion relations for the Hamilton-Jacobi functional for all $k<d/2$, but it has a pole at $k=d/2$, signifying the presence of a non-analytic (logarithmic) term in the solution. However, the divergent part of the full non-analytic solution is captured by the expansion in eigenfunctions of the dilatation operator, provided the pole at $k=d/2$ is regularized with the radial cutoff according to the dimensional regularization prescription \cite{Papadimitriou:2004ap,Papadimitriou:2011qb}
\be\label{DimReg}
\frac{1}{d-2k}\to \frac{r_c}{\ell} = -\frac12 \log e^{-2r_c/\ell}. 
\ee  
This prescription is equivalent to introducing a logarithmic term in the formal expansion of the Hamilton-Jacobi density from the start, as we did for generic AlAdS manifolds in \eqref{action_exp}.

For both even and odd dimensions, therefore, the general form of the boundary counterterms for asymptotically locally conformally flat AlAdS manifolds of arbitrary dimension is 
\be\label{ConFlatCounterterms}\boxed{
\cl\sbtx{ct}=-\sum_{k=0}^{[\frac d2]}\cl_{(2k)}=\frac{\sqrt{-h}}{\k^2}\Big(-\frac{d-1}{\ell}+\sum_{k=1}^{[\frac d2]}\frac{(-1)^{k}\ell^{2k-1}(2k-2)!(d-k)!}{2^{k-1}(k-1)!(d-2)!(d-2k)}\s_k(\bb P)\Big),}
\ee
with the regularization prescription \eqref{DimReg} understood in the case of even $d$. Curiously, these counterterms take the form of a Meissner-Olechowski theory of gravity \cite{Meissner:2000dy}, which is equivalent to Lovelock gravity for conformally flat metrics.

\subsection{Kounterterms for asymptotically conformally flat AlAdS manifolds} 

The form of the boundary Kounterterms \eqref{Kounterterms-density} for conformally flat manifolds can be deduced from their defining relations, given respectively in \eqref{Beven} and \eqref{c-even} for even boundary dimension $d$ and in \eqref{Bodd} and \eqref{c-odd} for odd $d$, by replacing the Riemann tensor of the induced metric with its conformally flat value. As we have seen in the previous subsection, there are two possible expressions for $\car_{ijkl}[h]$, following from either the third equation in \eqref{ConFlatEqs}, or the second equation in \eqref{ConFlatGeom}. These result in two different formulas for the Kounterterms on asymptotically conformally flat manifolds, both of which are useful for different purposes. 

\subsubsection*{Even dimensions} 

In section \ref{sec:Kounterterms}, we showed that the density $B_{2n-1}$ coincides with the pullback of the Chern form \eqref{CS-form} on the boundary of even dimensional AlAdS manifolds (see \eqref{Beven-CS}). Inserting the expression for the Riemann tensor following from the third equation in \eqref{ConFlatEqs} in the pullback of the Chern form in \eqref{pullbackPi}, we obtain
\bal
c_{2n-1}B_{2n-1}=&\;\frac{\sqrt{-h}}{\k^2}\frac{\sqrt{\p}(n-1)!}{2(2n-2)!}\sum_{k=0}^{n-1}\frac{(-\ell
^2)^{n-1-k}(2k)!(2n-2k-1)!}{k!\,\G\big(n-k+\frac12\big)}\s_{2n-2k-1}(\bb K)\NO\\
=&\;\frac{\sqrt{-h}}{\k^2}\frac{(n-1)!}{(2n-2)!}\sum_{k=0}^{n-1}\frac{(-4\ell
^2)^{n-k-1}(2k)!(n-k-1)!}{k!}\s_{2n-2k-1}(\bb K)\NO\\
=&\;\frac{\sqrt{-h}}{\k^2}\frac{(n-1)!}{(2n-2)!}\sum_{k=0}^{n-1}(-\ell
^2)^{k}\frac{2^{2k}k!(2n-2k-2)!}{(n-k-1)!}\s_{2k+1}(\bb K),
\eal
where the symmetric polynomials of the extrinsic curvature $\s_k(\bb K)$ were defined in \eqref{sigma-K}. Noticing that the $k=0$ term cancels the Gibbons-Hawking term in 
\eqref{Kounterterms-density}, we conclude that, for even dimensional asymptotically conformally flat AlAdS manifolds, the Kounterterms take the form
\be\label{ConFlatK-even-K}
\cl\sbtx{K}=\frac{\sqrt{-h}}{\k^2}\frac{(n-1)!}{(2n-2)!}\sum_{k=1}^{n-1}(-\ell
^2)^{k}\frac{2^{2k}k!(2n-2k-2)!}{(n-k-1)!}\s_{2k+1}(\bb K),\qquad d=2n-1.
\ee
Remarkably, this expression coincides with the boundary counterterms for such manifolds given in \eqref{Counterterms-ConFlat-even}, which correspond to an exact solution of the Hamilton-Jacobi equation.

The fact that the Kounterterms exactly coincide with the boundary counterterms for even dimensional asymptotically conformally flat AlAdS manifolds is intimately related with well known results in the mathematics literature \cite{Anastasiou:2018mfk}. As we reviewed above, the boundary integral of the density $B_{2n-1}$ can be expressed as a bulk integral over the Pfaffian of the bulk Riemann tensor, i.e. the bulk Euler density. For a conformally flat manifold $\cm_{2n}$, the Euler density is proportional to the symmetric polynomial $\s_n(P)$ of the corresponding (bulk) Schouten tensor (see e.g. Proposition 2.2 in \cite{10.1093/imrn/rnt095}). If $\cm_{2n}$ is also Einstein as in this case, the Schouten tensor is proportional to the metric (see \eqref{Schouten-on-shell}) and the Euler density reduces to a multiple of the bulk volume form. This is the content of Theorem 1.2 (see also Lemma 4.4) in \cite{albin2005renormalizing} and Theorem 3.6 in \cite{chang2005renormalized}, which generalize the result of Anderson for four dimensions \cite{Anderson_l2}. However, the volume of $\cm_{2n}$ is proportional to the on-shell action, which in turn coincides -- by construction -- with the boundary counterterms.  

In order to express the Kounterterms in terms of the Schouten tensor rather than the 
extrinsic curvature, one may insert in \eqref{ConFlatK-even-K} the expression for $K^i_j$ in terms of $\cp^i_j$ we determined in \eqref{KSol}. However, it is much simpler to start from an alternative expression for the Kounterterms for even dimensional conformally flat manifolds that is linear in the extrinsic curvature. Replacing the Riemann tensor of the induced metric in \eqref{Beven} using the second equation in \eqref{ConFlatGeom} leads to
\bal\label{ConFlatK-even-P}
c_{2n-1}B_{2n-1}=&\; \frac{\sqrt{-h}}{\k^2}\frac{(-\ell
^2)^{n-1}}{(2n-2)!}\int_{0}^{1}\tx dt\,\delta_{j_{1\ldots }j_{2n-1}}^{i_{1}\ldots
i_{2n-1}}K_{i_{1}}^{j_{1}}\Big((1-t^2)K_{i_{2}}^{j_{2}}K_{i_{3}}^{j_{3}}-\frac{1}{\ell^2}\d_{i_{2}}^{j_{2}}\d_{i_{3}}^{j_{3}}\Big) \times \notag \\
&\cdots \times \Big((1-t^2)K_{i_{2n-2}}^{j_{2n-2}}K_{i_{2n-1}}^{j_{2n-1}}-\frac{1}{\ell^2}\d_{i_{2n-2}}^{j_{2n-2}}\d_{i_{2n-1}}^{j_{2n-1}}\Big)\NO\\
=&\; \frac{\sqrt{-h}}{\k^2}\frac{(-\ell
^2)^{n-1}}{(2n-2)!}\int_{0}^{1}\tx dt\,\delta_{j_{1\ldots }j_{2n-1}}^{i_{1}\ldots
i_{2n-1}}K_{i_{1}}^{j_{1}}\Big(2(1-t^2)\d_{i_{2}}^{j_{2}}\cp_{i_{3}}^{j_{3}}-\frac{t^2}{\ell^2}\d_{i_{2}}^{j_{2}}\d_{i_{3}}^{j_{3}}\Big) \times \notag \\
&\cdots \times \Big(2(1-t^2)\d_{i_{2n-2}}^{j_{2n-2}}\cp_{i_{2n-1}}^{j_{2n-1}}-\frac{t^2}{\ell^2}\d_{i_{2n-2}}^{j_{2n-2}}\d_{i_{2n-1}}^{j_{2n-1}}\Big)\NO\\
=&\; \frac{\sqrt{-h}}{\k^2}\frac{[\G\big(\frac{d+1}{2}\big)]^2}{\G(d+1)\G(d)}\sum_{k=0}^{n-1}\frac{(-\ell^2)^{k}2^{3k}\G(d-2k)\G(d-k)}{[\G\big(\frac{d+1}{2}-k\big)]^2}\delta_{j_{1\ldots }j_{k+1}}^{i_{1}\ldots
i_{k+1}}K_{i_{1}}^{j_{1}}\cp^{i_2}_{j_2}\cdots \cp^{i_{k+1}}_{j_{k+1}},
\eal
where we have used the beta function integral
\be
\int_{0}^{1}\tx dt\,(1-t^2)^kt^{2n-2k-2}=\frac{2^{2k+1}k!n!(2n-2k-2)!}{(2n)!(n-k-1)!}.
\ee

While the form \eqref{ConFlatK-even-K} of the Kounterterms facilitates a direct comparison with the boundary counterterms, the expression \eqref{ConFlatK-even-P} will prove instrumental for comparing the Kounterterms for even and bulk dimensions. Moreover, it is particularly useful for determining the covariant asymptotic expansion of the Kounterterms in eigenfunctions of the dilatation operator, since successive powers of the Schouten tensor are increasingly subleading asymptotically. This should be contrasted with the powers of the extrinsic curvature in \eqref{ConFlatK-even-K}, all of which are $\co(1)$ asymptotically. 

Inserting the expansion \eqref{K-exp-generic} of the extrinsic curvature in \eqref{ConFlatK-even-P}, up to terms of negative dilatation weight (i.e. asymptotically vanishing), we get
\bal
c_{2n-1}B_{2n-1} =&\; \frac{\sqrt{-h}}{\k^2}\frac{[\G\big(\frac{d+1}{2}\big)]^2}{\G(d+1)\G(d)}\sum_{k=0}^{n-1}\frac{(-\ell^2)^{k}2^{3k}\G(d-2k)\G(d-k)k!}{[\G\big(\frac{d+1}{2}-k\big)]^2}\sum_{m=0}^{n-1} \tr\big(\bb K\sub{2m}\bb T_{k}(\bb P)\big)\\
=&\; \frac{\sqrt{-h}}{\k^2}\frac{[\G\big(\frac{d+1}{2}\big)]^2}{\G(d+1)\G(d)}\Big(\sum_{m=0}^{n-1}\sum_{k=0}^{m}\frac{(-\ell^2)^{k}2^{3k}\G(d-2k)\G(d-k)k!}{[\G\big(\frac{d+1}{2}-k\big)]^2}\tr\big(\bb K\sub{2m-2k}\bb T_k(\bb P)\big)\Big),\NO
\eal
where $\bb T_k(\bb P)$ was defined in eq.~\eqref{k-th-Newton-P}. Inserting the solution \eqref{KSol} for the extrinsic curvature in this expression, the sum over $k$ can be evaluated using the following two results:
\bal\label{id7}
&\tr\Big[\Big(\bb T_{m-k}-\frac{d-m+k}{d-1}\s_{m-k}\bb 1\Big)\bb T_k\Big]\NO\\
&\;\;\stackrel{\eqref{product-rule}}{=}\sum_{l=0}^{m-k-1}\tr\big(\s_l\bb T_{m-l}-\s_{m-l}\bb T_{l}\big)+\Big(1-\frac{d-m+k}{d-1}\Big)\s_{m-k}\tr\bb T_k\NO\\
&\;\;\;\;\,=\sum_{l=0}^{m-k-1}\tr\big(\s_l\bb T_{m-l}-\s_{m-l}\bb T_{l}\big)+\tr\Big(\s_{m-k}\bb T_k-\frac{d-k}{d-1}\s_k\bb T_{m-k}\Big)\NO\\
&\;\;\,\stackrel{\eqref{id5}}{=}-\frac{m!(d-m)!(d-2)!}{(m-k)!k!(d-k-1)!(d-m+k-1)!}\s_m,\qquad k<m,
\eal
and
\bal\label{id8}
&\frac{[\G\big(\frac{d+1}{2}\big)]^2}{\G(d+1)\G(d)}\Big(\sum_{k=0}^{m-1}\frac{2^{4k}\G(d-2k)}{[\G\big(\frac{d+1}{2}-k\big)]^2}\frac{(2m-2k-2)!}{(m-k)!(m-k-1)!}-\frac{2^{4m-1}\G(d-2m)}{[\G\big(\frac{d+1}{2}-m\big)]^2}\Big)\NO\\
&=\frac{(2m-2)!}{m!(m-1)!(d-1)!}\Big(1-\frac{d-1}{d-2m}\Big).
\eal

Putting everything together, we conclude that, in terms of the Schouten tensor of the induced metric, the Kounterterm density \eqref{Kounterterms-density} in even bulk dimensions is given by
\be\label{ConFlatK-even}\boxed{
\cl\sbtx{K}=\frac{\sqrt{-h}}{\k^2}\Big(-\frac{d-1}{\ell}+\sum_{k=1}^{[\frac d2]}\frac{(-1)^{k}\ell^{2k-1}(2k-2)!(d-k)!}{2^{k-1}(k-1)!(d-2)!(d-2k)}\s_k(\bb P)\Big),\qquad d=2n-1,}
\ee
in exact agreement with the boundary counterterms in \eqref{ConFlatCounterterms}. Of course, this result was expected since we have already shown above that the Kounterterms for even bulk dimensions coincide with the counterterms, but it verifies the equivalence of the two expressions for the Kounterterms.

\subsubsection*{Odd dimensions}

Analogous expressions for the Kounterterms can be derived for odd dimensional conformally flat AlAdS manifolds, but there are some key differences. Firstly, as for even dimensions, we can express the Kounterterms in terms of the extrinsic curvature of the induced metric by utilizing the third equation in \eqref{ConFlatEqs}. Inserting this in \eqref{Bodd} gives  
\bal\label{ConFlatK-odd-K}
c_{2n}B_{2n}=&\;\frac{\sqrt{-h}}{\k^2}\frac{(-\ell ^{2})^{n-1}}{%
2^{2n-2}(n-1)!^{2}}\int_{0}^{1}\tx dt\int_{0}^{t}\tx ds\,\d_{j_{1}\ldots j_{2n}}^{i_{1}\ldots i_{2n}}K_{i_{1}}^{j_{1}}\d_{i_{2}}^{j_{2}}\Big((1-t^{2})K_{i_{3}}^{j_{3}}K_{i_{4}}^{j_{4}}-(1-s^{2})
\frac{1}{\ell^2}\d_{i_{3}}^{j_{3}}\d_{i_{4}}^{j_{4}}\Big)
\times \NO \\
&\;\cdots \times \Big((1-t^{2})K_{i_{2n-1}}^{j_{2n-1}}K_{i_{2n}}^{j_{2n}}-(1-s^2)
\frac{1}{\ell^2}\d_{i_{2n-1}}^{j_{2n-1}}\d_{i_{2n}}^{j_{2n}}\Big)\\
=&\;\frac{\sqrt{-h }}{\k^2}\sum_{k=0}^{n-1}\int_{0}^{1}\tx dt\,(1-t^2)^{k}\int_{0}^{t}\tx ds\,(1-s^2)^{n-k-1}\frac{(-\ell^2)^{k}(2n-2k-1)!(2k+1)!}{2^{2n-2}(n-1)!k!(n-k-1)!}\s_{2k+1}(\bb K)\NO\\
=&\;\frac{\sqrt{-h }}{\k^2}\sum_{k=0}^{n-1}\frac{(-\ell^2)^{k}(2n-2k-1)!(2k+1)!}{2^{2n-1}(n-1)!(k+1)!(n-k-1)!}{}_3F_2\Big(\frac12,1,1+k-n;\frac32,k+2;1\Big)\s_{2k+1}(\bb K),\NO
\eal
where we have used the integral representation of the generalized hypergeometric function
\be
\int_{0}^{1}\tx dt\,(1-t^2)^{k}\int_{0}^{t}\tx ds\,(1-s^2)^{n-k-1}=\frac{1}{2(k+1)}\;{}_3F_2\Big(\frac12,1,1+k-n;\frac32,k+2;1\Big).
\ee

It can be easily verified that this expression (after subtracting the Gibbons-Hawking term) does not satisfy the Hamilton-Jacobi equation \eqref{HJdensity-eq-sigma}. In fact, this is guaranteed by our observation earlier that, for odd bulk dimensions, any solution of the Hamilton-Jacobi equation that satisfies the condition \eqref{HJbc} cannot be polynomial in the extrinsic curvature. An immediate consequence is that, for odd dimensions, the Kounterterms do not generically agree with the boundary counterterms, even in the special case of conformally flat manifolds. This confirms our earlier conclusion, drawn by comparing the general expansions \eqref{Kounterterms-R-low-d} and \eqref{counterterms-low-d}. Nevertheless, the identity we derive next shows that, for odd dimensional conformally flat AlAdS manifolds, the Kounterterms deviate from the boundary counterterms in a very specific way that can be quantified. 

Using the second equation in \eqref{ConFlatGeom} to replace the Riemann tensor in \eqref{Beven} gives instead 
\bal\label{ConFlatK-odd-P}
c_{2n}B_{2n}=&\;\frac{\sqrt{-h}}{\k^2}\frac{(-\ell ^{2})^{n-1}}{%
2^{2n-2}(n-1)!^{2}}\int_{0}^{1}\tx dt\int_{0}^{t}\tx ds\,\d_{j_{1}\ldots j_{2n}}^{i_{1}\ldots i_{2n}}K_{i_{1}}^{j_{1}}\d_{i_{2}}^{j_{2}}\Big(2(1-t^{2})\d_{i_{3}}^{j_{3}}\cp_{i_{4}}^{j_{4}}+(s^{2}-t^2)
\frac{1}{\ell^2}\d_{i_{3}}^{j_{3}}\d_{i_{4}}^{j_{4}}\Big)
\times \notag \\
&\;\cdots \times \Big(2(1-t^{2})\d_{i_{2n-1}}^{j_{2n-1}}\cp_{i_{2n}}^{j_{2n}}+(s^2-t^2)
\frac{1}{\ell^2}\d_{i_{2n-1}}^{j_{2n-1}}\d_{i_{2n}}^{j_{2n}}\Big)\NO\\
=&\;\frac{\sqrt{-h}}{\k^2}\sum_{k=0}^{n-1}\frac{(-\ell^2)^{k}2^{-k}\G(d-k)\G\big(\frac{d}{2}-k+1\big)\G\big(\frac{d}{2}-k\big)}{\G\big(\frac{d}{2}+1\big)\G\big(\frac{d}{2}\big)(d-2k)!}\d_{j_{1}\ldots j_{k+1}}^{i_{1}\ldots i_{k+1}}K_{i_{1}}^{j_{1}}\cp_{i_{2}}^{j_{2}}\cdots\cp_{i_{k+1}}^{j_{k+1}},
\eal
where we have used the identity
\be
\int_{0}^{1}\tx dt\int_{0}^{t}\tx ds\,(1-t^2)^k(t^2-s^2)^{n-k-1}=\frac{2^{2n-2k-2}k!(n-k)![(n-k-1)!]^2}{n!(2n-2k)!}.
\ee
Remarkably, \eqref{ConFlatK-odd-P} coincides with \eqref{ConFlatK-even-P} when both are expressed in terms of the boundary dimension $d$ (which is of course different in the two cases), except for the upper limit in the summation over $k$. This provides a proof that the Kounterterms for even and odd dimensions are identical as functions of the boundary dimension $d$, except for the finite terms arising for odd bulk dimensions.

Finally, the result \eqref{ConFlatK-odd-P} allows us to pinpoint the difference between the Kounterterms and the boundary counterterms for odd dimensional asymptotically conformally flat AlAdS manifolds. We have already shown that the boundary counterterms coincide with the Kounterterms in even dimensions and that the Kounterterms for even and odd dimensions coincide when expressed in terms of the boundary dimension $d$, except for finite terms. It follows that the Kounterterms for odd dimensional conformally flat AlAdS manifolds differ from the boundary counterterms only by logarithmic and local finite terms. 

To quantify this difference, we insert the covariant expansion of the extrinsic curvature in \eqref{ConFlatK-odd-P} and drop all terms of negative dilatation weight: 
\bal\label{ConFlatK-odd-Pexp}
c_{2n}B_{2n} =&\; \frac{\sqrt{-h}}{\k^2}\frac{[\G\big(\frac{d+1}{2}\big)]^2}{\G(d+1)\G(d)}\sum_{k=0}^{n-1}\frac{(-\ell^2)^{k}2^{3k}\G(d-2k)\G(d-k)k!}{[\G\big(\frac{d+1}{2}-k\big)]^2}\sum_{m=0}^{n} \tr\big(\bb K\sub{2m}\bb T_{k}(\bb P)\big)\NO\\
=&\; \frac{\sqrt{-h}}{\k^2}\frac{[\G\big(\frac{d+1}{2}\big)]^2}{\G(d+1)\G(d)}\Big(\sum_{m=0}^{n-1}\sum_{k=0}^{m}\frac{(-\ell^2)^{k}2^{3k}\G(d-2k)\G(d-k)k!}{[\G\big(\frac{d+1}{2}-k\big)]^2}\tr\big(\bb K\sub{2m-2k}\bb T_k(\bb P)\big)\NO\\
&+\sum_{k=0}^{n-1}\frac{(-\ell^2)^{k}2^{3k}\G(d-2k)\G(d-k)k!}{[\G\big(\frac{d+1}{2}-k\big)]^2}\tr\big(\bb K\sub{2n-2k}\bb T_k(\bb P)\big)\Big).
\eal
We have deliberately expressed all coefficients in terms of the boundary dimension and have used the parameterization arising naturally in even bulk dimensions (see \eqref{ConFlatK-even-P}). This renders $c_d B_d$ for even and odd bulk dimensions manifestly identical, except for the sum in the last line of \eqref{ConFlatK-odd-Pexp}. 

Evaluating the sums in \eqref{ConFlatK-odd-Pexp} using identities \eqref{id7} and \eqref{id8}, we find that the Kounterterm density \eqref{Kounterterms-density} for odd dimensional asymptotically conformally flat AlAdS manifolds takes the form 
\bal
\cl\sbtx{K}=&\;\frac{\sqrt{-h}}{\k^2}\Big(-\frac{d-1}{\ell}+\sum_{k=1}^{[\frac d2]-1}\frac{(-1)^{k}\ell^{2k-1}(2k-2)!(d-k)!}{2^{k-1}(k-1)!(d-2)!(d-2k)}\s_k(\bb P)\\
&+\frac{(-1)^{n}\ell^{2n-1}(d-n)!}{2^{n-1}}\Big(\frac{(2n-2)!}{(n-1)!(d-2)!(d-2n)}-\frac{2^{4n-1}[\G\big(\frac{d+1}{2}\big)]^2\G(d-2n)n!}{\G(d+1)\G(d)[\G\big(\frac{d+1}{2}-n\big)]^2}\Big)\s_n(\bb P)\Big).\NO
\eal
Notice that both terms in the last line of this expression have simple poles at $n=d/2$. However, in contrast to the boundary counterterms \eqref{ConFlatCounterterms} that solve the Hamilton-Jacobi equation, the poles in the Kounterterms cancel to produce a finite result, namely  
\begin{align}
\label{ConFlatK-odd}
\boxed{
\begin{aligned}
\cl\sbtx{K}=&\;\frac{\sqrt{-h}}{\k^2}\Big(-\frac{d-1}{\ell}+\sum_{k=1}^{[\frac d2]-1}\frac{(-1)^{k}\ell^{2k-1}(2k-2)!(d-k)!}{2^{k-1}(k-1)!(d-2)!(d-2k)}\s_k(\bb P)\\
&+\frac{(-1)^{\frac d2}\ell^{d-1}}{2^{\frac d2}}\Big(1+\frac{2d}{d-1}-2d\log 2-d\j\Big(\frac{d+1}{2}\Big)+d\j(d-1)\Big)\s_{\frac d2}(\bb P)\Big),\quad d=2n,
\end{aligned}
}
\end{align}
where $\j(z)=\G'(z)/\G(z)$ is the digamma function. It can be easily checked that the last line in this expression reproduces all finite terms in \eqref{Kounterterms-R-low-d} upon setting the Weyl tensor there to zero. This result pinpoints the reason why the Kounterterms fail to capture the logarithmic divergence of odd dimensional asymptotically conformally flat AlAdS manifolds and provides a general expression for the local finite term that the Kounterterms produce instead.

\section{Discussion}
\label{disc}

We have identified necessary and sufficient conditions for the Kounterterms to regularize the AdS variational problem in arbitrary dimension. A well posed variational problem for AlAdS manifolds exists only within the space of asymptotically Einstein manifolds. Within this space, the extrinsic curvature and the induced metric of the boundary are asymptotically related, permitting a direct comparison between the Kounterterms and the boundary counterterms obtained via holographic renormalization. Comparison for dimensions three to seven shows that, except in four dimensions, a necessary condition for agreement is the vanishing of the boundary Weyl tensor. 

By determining the general form of the boundary counterterms for AlAdS manifolds of arbitrary dimension with zero boundary Weyl tensor, we showed that in even bulk dimensions the vanishing of the boundary Weyl tensor is also a sufficient condition for agreement between the Kounterterms and the boundary counterterms. However, this is not a sufficient condition in odd bulk dimensions. The disagreement in that case arises solely from the logarithmic divergence related to the holographic conformal anomaly. In particular, the boundary counterterms contain a logarithmically divergent term proportional to the holographic conformal anomaly, or Branson's $Q$-curvature, as it is known in the mathematics literature. For AlAdS manifolds with vanishing boundary Weyl tensor, this quantity is proportional to the determinant of the boundary Schouten tensor, $\det\bb P$, which in this case coincides with the Euler-Poincar\'e density of the boundary. In contrast, the logarithmically divergent term is absent in the Kounterterms, but a finite term proportional to $\det\bb P$ arises instead. It follows that necessary and sufficient conditions for the Kounterterms to regularize the AdS variational problem in odd dimensions are (a) zero boundary Weyl tensor and (b) zero boundary Euler characteristic. These conditions are summarized in table~\ref{conditions}.     

Although the Kounterterms agree with the boundary counterterms at the level of the action once the above conditions are met, it is not guaranteed that their respective contributions to the quasilocal stress tensor, or to higher order moments (i.e. holographic correlation functions), also agree. In section \ref{sec:conflat} we showed that agreement at the level of the action implies agreement of the corresponding radial canonical momenta, i.e. of the corresponding quasilocal stress tensors. This ensures that the Kounterterms correctly renormalize the conserved charges of AlAdS black holes with a conformally flat boundary and --in the case of odd dimensions-- zero boundary Euler characteristic. We anticipate that, under these conditions, higher order holographic correlation functions may or may not be renormalized by the Kounterterms, depending on the spacetime dimension and the order of the correlation function. It would be interesting to address this question.   

It would also be interesting to generalize our analysis to other theories of gravity that admit AdS solutions for which a version of Kounterterms exists, such as higher derivative theories \cite{Giribet:2018hck} and holographic entanglement entropy \cite{Anastasiou:2018rla,Anastasiou:2019ldc}. As for pure Einstein-Hilbert gravity, we anticipate that, in general, the Kounterterms cancel the long distance divergences and regularize the variational problem only for solutions with a conformally flat boundary.

\section*{Acknowledgments}
IP thanks the Pontificia Universidad Católica de Valparaíso and the Universidad Andrés Bello for hospitality and partial financial support during the early stages of this project. This work was funded in part by FONDECYT Grants No.~3190314 (GA), No.~1190533 (OM) and No.~1170765 (RO), as well as VRIEA-PUCV Grant No.~123.764/2019. IP is supported by a KIAS Individual Grant
(PG064402) at the Korea Institute for Advanced Study. We also acknowledge extensive use of the Mathematica package \href{http://www.xact.es/}{xAct}.

\appendix

\section*{Appendices}

\appendix{}

\section{Radial foliation of AlAdS manifolds}  
\label{RadialFoliation}

As was reviewed in section \ref{sec:hr}, using a suitable Gaussian normal coordinate, $r$, in an open neighborhood of the conformal boundary, $\pa\cm$, of a $d+1$ dimensional AlAdS manifold, $\cm$, the metric $g$ on $\cm$ can be written in the Fefferman-Graham form (see eq.~\eqref{FG-canonical}) 
\be\label{FG-appendix}
g=\tx dr^2+h_{ij}(r,x)\tx dx^i\tx dx^j,\qquad i,j=1,\ldots d,
\ee
where $h_{ij}$ is the induced metric on a radial slice $\S_r\cong \pa\cm$, and the asymptotic boundary is located at $r\to\infty$. In this appendix, we compile several identities that express the intrinsic curvature of $\cm$ in terms of the intrinsic and extrinsic curvatures of the radial slices $\S_r$. 

Our notation throughout this article is as follows. A dot $\dot{}$ stands for the radial derivative $\pa_r$, $D_i$ denotes the covariant derivative with respect to the induced metric $h_{ij}$, while $K_{ij}=\frac12\dot h_{ij}$ is the extrinsic curvature of $\S_r$ in $\cm$. Moreover, we use the convention that the radial derivative is applied after any indices are raised or lowered with $h_{ij}$ and its inverse. For example, $\dot K^i_j=\pa_r K^i_j\neq h^{ik}\dot K_{kj}$. Finally, script symbols such as $\car_{ij}$ or $\cp_{ij}$ denote curvature tensors of the induced metric $h_{ij}$.

\subsection{Off-shell identities}

We begin with the identities that follow solely from the decomposition \eqref{FG-appendix} of the bulk metric $g$, without imposing Einstein's equations \eqref{einstein}.

\paragraph{Christoffel symbols} The only non vanishing components of the Christoffel symbols of $g$ are
\be
\G^r_{ij}[g]=-K_{ij},\qquad \G^i_{rj}[g]=K^i_j,\qquad \G^i_{jk}[g]=\G^i_{jk}[h].
\ee

\paragraph{Riemann tensor} The Riemann tensor of $g$ decomposes as 
\bea\label{Riemann_components}
&&R^i{}_{rjr}
=-\dot{K}^i_j-K^i_kK^k_j,\NO\\
&& R^i{}_{kjr}
=D^iK_{kj}-D_kK^i_j,\NO\\
&&R^{ik}{}_{jl}=
\car^{ik}{}_{jl}-K^i_jK^k_l+K^i_lK^k_j.
\eea

\paragraph{Ricci tensor} From \eqref{Riemann_components} it follows that the components of the Ricci tensor take the form 
\bea\label{Ricci_components}
&&R_{rr}=-\dot K-K^k_lK^l_k,\NO\\
&&R_{ri}=D_j K^j_i-D_i K,\NO\\
&&R^i_j=\car^i_j-KK^i_j-\dot K^i_j.
\eea

\paragraph{Ricci scalar} These in turn determine the Ricci scalar, which is given by
\be
R=\car -K^2-2\dot K-K^k_lK^l_k.
\ee

\paragraph{Weyl tensor} The definition \eqref{BWeyl} of the bulk Weyl tensor implies that 
\bal\label{Weyl_components_off-shell}
X^i_j\equiv &\; W^i{}_{rjr} &&\hskip-3.9cm= R^i{}_{rjr}-\d^i_j P_{rr}-P^i_j,\NO\\
Y^i{}_{kj}\equiv &\;  W^i{}_{kjr} &&\hskip-3.9cm= R^i{}_{kjr}+h_{kj}P^i_r-\d^i_jP_{rk},\NO\\
Z^{ik}{}_{jl}\equiv &\; W^{ik}{}_{jl} &&\hskip-3.9cm= R^{ik}{}_{jl}+\d^i_l P^k_j+\d^k_j P^i_l-\d^i_j P^k_l-\d^k_l P^i_j,
\eal 
where we have introduced the symbols $X^i_j$, $Y^i{}_{kj}$ and $Z^{ik}{}_{jl}$ for the components of the Weyl tensor for later convenience. Notice that $X^i_j$ corresponds to the `electric part' of the Weyl tensor and is related to the components $Z^{ik}{}_{jl}$ through the trace identity $W^\m{}_{\r\m\s}=0$, which reads
\be\label{Weyl-trace}
Z^{ik}{}_{jk}+X^i_j=0.
\ee
This relation can be checked using \eqref{Weyl_components_off-shell}. Explicit expressions for the Weyl tensor components in terms of the intrinsic and extrinsic curvatures of $h_{ij}$ can be easily obtained using \eqref{Weyl_components_off-shell} and the above decomposition of the Riemann and Ricci curvatures. 

\subsection{On-shell identities}

Next, we collect a number of identities following not only from the radial foliation \eqref{FG-appendix}, but also from imposing Einstein's equations \eqref{einstein}. These imply that the bulk Ricci scalar is constant on-shell
\be\label{RicchiScalar-on-shell}
R=-\frac{d(d+1)}{\ell^2},
\ee 
while the Schouten tensor takes the form
\be\label{Schouten-on-shell}
P_{\m\n}=-\frac{1}{2\ell^2}g_{\m\n}.
\ee

\paragraph{Gauss-Codazzi equations} Einstein's equations decompose into the three equations
\bal\label{Gauss-Codazzi} 
&K^2-K^i_jK^j_i=\car+\frac{d(d-1)}{\ell^2},\NO\\
&D_j K^j_i-D_i K=0,\NO\\
&\car^i_j-KK^i_j-\dot K^i_j+\frac{d}{\ell^2}\d^i_j=0.
\eal

\paragraph{Weyl tensor} On-shell, the components of the Weyl tensor defined in \eqref{Weyl_components_off-shell} become
\bal\label{Weyl-on-shell}
X^i_j=&\; -\dot{K}^i_j-K^i_kK^k_j+\frac{1}{\ell^2}\d^i_j=-\car^i_j+KK^i_j-K^i_kK^k_j-\frac{d-1}{\ell^2}\d^i_j,\NO\\
Y^i{}_{kj}=&\; D^iK_{kj}-D_kK^i_j,\NO\\
Z^{ik}{}_{jl}
=&\;\car^{ik}{}_{jl}-K^i_jK^k_l+K^i_lK^k_j-\frac{1}{\ell^2}\d^i_l \d^k_j+\frac{1}{\ell^2}\d^i_j \d^k_l.
\eal
Besides the trace identity \eqref{Weyl-trace} that holds off-shell, the Gauss-Codazzi equations \eqref{Gauss-Codazzi} imply that 
\be
Z^{ij}{}_{ij}=-X^i_i=0,\qquad Y^i{}_{ji}=0.
\ee

\paragraph{Flow equations for the Weyl tensor} The definition \eqref{BWeyl}, together with the on-shell expression for the Schouten tensor \eqref{Schouten-on-shell}, imply that, on-shell, the Weyl tensor satisfies the Bianchi identity
\be
\nabla_\l W^\m{}_{\n\r\s}+\nabla_\r W^\m{}_{\n\s\l}+
\nabla_\s W^\m{}_{\n\l\r}=0.
\ee
Decomposing this into radial and transverse components leads to the three equations
\bal\label{Bianchi}
&\dot{Y}_{ij}{}^k=-D_iX^k_j+D_jX^k_i-2 K_{[i}^lY_l{}^k
{}_{j]}- K^k_lY_{ij}{}^l,\NO\\
&\dot{Z}^{ik}{}_{jl}= K^i_pZ^{kp}{}_{jl}- K^k_pZ^{ip}
{}_{jl}+D_jY^{ki}{}_l-D_lY^{ki}{}_j+ K^i_jX^k_l-
 K^i_lX^k_j+X^i_jK^k_l-X^i_lK^k_j,\NO\\
&D_p Z^{ik}{}_{jl}+D_j Z^{ik}{}_{lp}+D_l Z^{ik}{}_{pj}
=K^i_jY_{lp}{}^k+K^i_lY_{pj}{}^k+K^i_pY_{jl}{}^k-
K^k_jY_{lp}{}^i-K^k_lY_{pj}{}^i-K^k_pY_{jl}{}^i.
\eal

The first two equations in \eqref{Bianchi} correspond to geometric flow equations for the components ${Y}_{ij}{}^k$ and ${Z}^{ik}{}_{jl}$ of the bulk Weyl tensor. Recall that the components $X^i_j$ are not independent due to the trace identity \eqref{Weyl-trace}. Together with the trace of the flow equation for ${Z}^{ik}{}_{jl}$, \eqref{Weyl-trace} leads to the flow equation for the electric part of the Weyl tensor
\be
\dot{X}^{i}_{j}+KX^i_j-K^k_jX^i_k=D_kY^{ki}{}_j+K^k_lZ^{il}{}_{jk},
\ee
or equivalently
\be
\dot{X}^{i}_{j}+KX^i_j=D_kY^{ki}{}_j+\Big(K^k_l-\frac1\ell \d^k_l\Big)Z^{il}{}_{jk}+\Big(K^k_j-\frac1\ell \d^k_j\Big)X^i_k.
\ee

\subsection{Asymptotic expansions}

Using the first few orders in the covariant expansion of the canonical momentum $\p^{ij}$ given in \eqref{General-Pis}, one finds that the covariant asymptotic expansion of the extrinsic curvature \eqref{K-exp-generic} takes the form 
\be\label{K-exp}
K^i_j=\frac{1}{\ell}\d^i_j+\ell\,\cp^i_j+\frac{\ell^3}{(d-4)(d-2)}\Big[\cb^i_j+(d-4)\Big(\cp^{ik}\cp_{kj}-\cp\cp^i_j-\frac{1}{2(d-1)}\big(\cp^{kl}\cp_{kl}-\cp^2\big)\d^i_j\Big)\Big]+\cdots,
\ee
where the Bach tensor, $\cb^i_j$, is defined in \eqref{bBach}. Inserting this covariant expansion in the expressions for the components of the bulk Weyl tensor in \eqref{Weyl-on-shell}, we find that their leading asymptotic form coincides respectively with the Bach, Cotton and Weyl tensors of the induced metric $h_{ij}$, namely 
\be\label{Weyl-asymptotics}
X^i_j=\frac{\ell^2}{d-4}\cb^i_j+\cdots,\qquad Y^i{}_{kj}=\ell\, \cc_{jk}{}^i+\cdots,\qquad Z^{ik}{}_{jl}=\cw^{ik}{}_{jl}+\cdots,
\ee
where the ellipses denote terms with higher dilatation weight.

\section{Symmetric polynomials}
\label{SymmTr}

In this appendix we collect a few elementary properties of symmetric polynomials that we use in the main text. A discussion of symmetric polynomials in the context of conformal geometry can be found in \cite{10.1093/imrn/rnt095} and references therein.

Given a $d\times d$ matrix $\bb M$ with components $(\bb M)^i_j=M^i_j$, we define its $k$-th symmetric polynomial  
\be
\s_k(\bb M)\equiv\frac{1}{(d-k)!k!}\d^{j_1j_2\cdots j_d}_{i_1i_2\cdots i_d}M^{i_1}_{j_1}M^{i_2}_{j_2}\cdots M^{i_k}_{j_k}\d^{i_{k+1}}_{j_{k+1}}\cdots \d^{i_{d}}_{j_{d}}=\frac{1}{k!}\d^{j_1j_2\cdots j_k}_{i_1i_2\cdots i_k}M^{i_1}_{j_1}M^{i_2}_{j_2}\cdots M^{i_k}_{j_k},
\ee
where the generalized Kronecker delta was defined in \eqref{antisym-deltas}. Notice that $\s_{k}(\bb M)=0$ for $k>d$. The generating function of these polynomials is the determinant  
\be\label{det}
f(t)\equiv\det(\bb 1+t \bb M)=\exp\big(\tr\log(\bb 1+t\bb M)\big)=\sum_{k=0}^dt^k \s_k(\bb M).
\ee
This is a special case of the more general identity
\be
\s_k(\bb 1+t\bb M)=\sum_{m=0}^k\frac{(d-m)!}{(d-k)!(k-m)!}t^m\s_m(\bb M).
\ee
In particular,
\be
\s_k(\bb 1)=\left(\begin{matrix} d\\ k\end{matrix}\right)=\frac{d!}{k!(d-k)!}.
\ee

The generating function \eqref{det} leads to an alternative representation of the symmetric polynomials in terms of traces of powers of the matrix $\bb M$. Namely,  
\bal
\s_0(\bb M)=&\;1,\NO\\
\s_1(\bb M)=&\;\tr \bb M,\NO\\
\s_2(\bb M)=&\;\frac12\big((\tr \bb M)^2-\tr(\bb M^2)\big),\NO\\
\s_3(\bb M)=&\;\frac16\big((\tr \bb M)^3-3\tr\bb M\,\tr(\bb M^2)+2\tr(\bb M^3)\big),\NO\\
\s_4(\bb M)=&\;\frac{1}{24}\big((\tr \bb M)^4-6(\tr\bb M)^2\tr(\bb M^2)+3(\tr(\bb M^2))^2+8\tr\bb M\,\tr(\bb M^3)-6\tr(\bb M^4)\big),\NO\\
\vdots\,&\NO\\
\s_d(\bb M)=&\;\det \bb M.
\eal

A related object that plays an important role in our analysis is the $k$-th Newton transform of the matrix $\bb M$, which is defined as 
\be
(\bb T_k(\bb M))^i_j\equiv\frac{\pa}{\pa M^j_i}\s_{k+1}(\bb M)=\frac{1}{k!}\d^{ii_2\cdots i_{k+1}}_{jj_2\cdots j_{k+1}}M^{j_2}_{i_2}\cdots M^{j_{k+1}}_{i_{k+1}}=\sum_{m=0}^k(-1)^m\s_{k-m}(\bb M)(\bb M^m)^i_j.
\ee
The trace of the $k$-th Newton transform is proportional to $\s_k$, namely
\be
\tr\bb T_k(\bb M)=(d-k)\s_k(\bb M).
\ee

The generating function of these matrix-valued polynomials follows directly from  \eqref{det}:
\be
\frac{\pa}{\pa M^j_i}\det(\bb 1+t \bb M)=\det(\bb 1+t \bb M)\frac{\pa}{\pa M^j_i}\tr\log(\bb 1+t\bb M)=t\det(\bb 1+t \bb M)\Big(\frac{\bb 1}{\bb 1+t\bb M}\Big)^i_j=\sum_{k=1}^dt^k (\bb T_{k-1})^i_j,
\ee
and hence
\be
\bb F(t)\equiv\det(\bb 1+t \bb M)(\bb 1+t\bb M)^{-1}=\sum_{k=0}^{d-1}t^k \bb T_{k}(\bb M).
\ee
Notice that 
\be
\bb T_d(\bb M)=0,
\ee
by virtue of the Cayley-Hamilton theorem.

The $n$-th derivative of the generating function $\bb F(t)$ with respect to the parameter $t$ is given by
\be
\bb F^{(n)}(t)=\big(f^{(n)}(t)\bb 1-n\bb M\bb F^{(n-1)}(t)\big)(\bb 1+t\bb M)^{-1},
\ee
where $f^{(n)}(t)$ is the $n$-th derivative of the determinant \eqref{det}.
Evaluating this identity at $t=0$ leads to the recursion relation
\be\label{T-id}
\bb T_k(\bb M)=\s_k(\bb M)\bb 1-\bb M\bb T_{k-1}(\bb M),\qquad \bb T_0(\bb M)=\bb 1.
\ee

Finally, the following product rule can be easily proved by induction
\be\label{product-rule}
\bb T_m(\bb M)\bb T_n(\bb M) = \sum_{k=0}^m\s_k(\bb M)\bb T_{m+n-k}(\bb M)-\sum_{k=0}^{m-1}\s_{m+n-k}(\bb M)\bb T_{k}(\bb M).
\ee


\bibliographystyle{JHEP}
\bibliography{refs}

\end{document}